\documentclass{jpp}
\usepackage{graphicx}
\usepackage{natbib}
\usepackage[T1]{fontenc}
\usepackage{bm}
\usepackage{color}
\usepackage{graphicx}
\usepackage{amsmath}
\usepackage{amssymb}
\usepackage{amsfonts}
\usepackage{courier}
\usepackage{txfonts}
\ifCUPmtlplainloaded \else
  \checkfont{eurm10}
  \iffontfound
    \IfFileExists{upmath.sty}
      {\typeout{^^JFound AMS Euler Roman fonts on the system,
                   using the 'upmath' package.^^J}%
       \usepackage{upmath}}
      {\typeout{^^JFound AMS Euler Roman fonts on the system, but you
                   dont seem to have the}%
       \typeout{'upmath' package installed. JPP.cls can take advantage
                 of these fonts, if you use 'upmath' package.^^J}%
      }
  \else
  \fi
\fi

\ifCUPmtlplainloaded \else
  \checkfont{msam10}
  \iffontfound
    \IfFileExists{amssymb.sty}
      {\typeout{^^JFound AMS Symbol fonts on the system, using the
                'amssymb' package.^^J}%
       \usepackage{amssymb}%
         
         \let\geq=\geqslant
      }{}
  \fi
\fi

\ifCUPmtlplainloaded \else
  \IfFileExists{amsbsy.sty}
    {\typeout{^^JFound the 'amsbsy' package on the system, using it.^^J}%
     \usepackage{amsbsy}}
    {\providecommand\boldsymbol[1]{\mbox{\boldmath $##1$}}}
\fi

\newcommand{\unit}[1]{\ensuremath{\bm{\widehat{\mathrm{#1}}}}}
\usepackage[T1]{fontenc}
\usepackage{bm}
\usepackage{color}
\usepackage{graphicx}

\usepackage{ae}
\usepackage{amsmath}
\usepackage{amssymb}
\usepackage{amsfonts}
\usepackage{mathrsfs}
\usepackage{units}
\usepackage{natbib}
\usepackage[disable]{todonotes}

\newcommand{\be}{\begin{displaymath}}
\newcommand{\ee}{\end{displaymath}}
\newcommand{\bn}{\begin{equation}}
\newcommand{\en}{\end{equation}}

\newcommand{\lpa}{\left(}
\newcommand{\rpa}{\right)}

\usepackage{enumerate}

\newcommand{\dd}{\mathrm{d}}
\newcommand{\nofrac}[2]{#1/#2}
\newcommand{\tallrow}{\\[-0.8em]}
\newcommand{\zkba}{B--A}
\newcommand{\Eceffeq}{E_{\rm c}^{\rm eff}}
\newcommand{\Ectot}{E_{\rm c}^{\rm tot}}

\newcommand{\Eceff}{$E_{\rm c}^{\rm eff}$}
\newcommand{\taur}{\tau_{\rm syn}}

\newcommand{\lnLStar}{\ln\Lambda_{\rm c}}

\newcommand{\Fbr}{F_{\rm br}}

\usepackage[T1]{fontenc}
\usepackage{bm}
\usepackage{color}
\usepackage{graphicx}
\usepackage[breaklinks=true]{hyperref}
\hypersetup{
  unicode=false,          % non-Latin characters in Acrobat’s bookmarks
  pdftoolbar=true,        % show Acrobat’s toolbar?
  pdfmenubar=true,        % show Acrobat’s menu?
  pdffitwindow=false,     % window fit to page when opened
  pdfstartview={FitH},    % fits the width of the page to the window
  pdftitle={Generalized collision operator for fast electrons interacting with
  partially ionized impurities},    % title
  pdfauthor={Hesslow, Embréus, Hoppe, DuBois, Papp, Rahm, and Fülöp},     % author
  pdfsubject={Subject},   % subject of the document
  pdfcreator={Creator},   % creator of the document
  pdfproducer={Producer}, % producer of the document
  pdfkeywords={keyword1} {key2} {key3}, % list of keywords
  pdfnewwindow=true,      % links in new PDF window
  colorlinks=true,        % false: boxed links; true: colored links
  linkcolor=blue,         % color of internal links (change box color with linkbordercolor)
  citecolor=blue,         % color of links to bibliography
  filecolor=blue,         % color of file links
  urlcolor=blue           % color of external links
}

\title[Fast-electron dynamics in partially ionized plasmas]{Generalized collision operator for fast electrons interacting with
  partially ionized impurities} 
\author{L.~Hesslow\aff{1}
  \corresp{\email{hesslow@chalmers.se}},  O.~Embr\'eus\aff{1}, M.~Hoppe\aff{1}, T.~C.~DuBois\aff{1}, G.~Papp\aff{2}, M.~Rahm\aff{3}, T.~F\"ul\"op\aff{1}}
\affiliation{\aff{1}Department of Physics, Chalmers University of Technology,
 SE-41296 G\"{o}teborg, Sweden
 \aff{2} Max Planck Institute for Plasma Physics, D-85748 Garching, Germany
 \aff{3} Department of Chemistry and Chemical Engineering, Chalmers University of Technology,
  SE-41296 Gothenburg, Sweden} 

\begin{document}
\maketitle
\begin{abstract}
  Accurate modelling of the interaction between fast electrons and
  partially ionized atoms is important for evaluating tokamak
  disruption mitigation schemes based on material injection.  This
  requires accounting for the effect of screening of the impurity
  nuclei by the cloud of bound electrons.  In this paper, we generalize the Fokker--Planck operator in a fully ionized plasma by accounting for the effect of screening.
  We detail the derivation of this generalized operator, and calculate the effective ion 
  length-scales, which are needed in the components of the collision
  operator, for a number of ion species commonly appearing  in fusion
  experiments.  We show that for high electric fields, the secondary
  runaway growth rate can be substantially larger than in a fully
  ionized plasma with the same effective charge, although the growth
  rate is significantly reduced at near-critical electric fields.
  Furthermore, by comparison with the Boltzmann collision operator, we
  show that the Fokker--Planck formalism is accurate even for large
  impurity content.
  \end{abstract}

\maketitle

%%%%%%%%%%%%%%%%%%%%%%%%%%%%%%%%%%%%%%%%%%%%%%%%%%%%%%%%%%%
\renewcommand{\thesection}{\Roman{section}}
\renewcommand{\thesubsection}{\Alph{subsection}} 

\section{Introduction}
Runaway acceleration of an electron in a plasma occurs if the electric
field exceeds a critical value, above which the friction force on the
electron from collisions with other plasma particles becomes smaller
than the force from the electric field \citep{Wilson1925}.  Electrons
can enter the runaway region in velocity space as a result of a random
walk caused by long-range Coulomb collisions (primary or Dreicer
generation) \citep{dreicer1959}. If there is an initial population of fast
electrons in the plasma, they may produce secondary runaway electrons
via close collisions -- leading to an exponential multiplication of the
fast electron population -- an avalanche
\citep{sokolov1979multiplication}.  Secondary generation of runaway
electrons is expected to be substantial in future high-current tokamak
disruptions \citep{jayakumar1993,RosenbluthPutvinski1997}, and
successful mitigation is required to prevent unacceptable wall damage
if a runaway population is formed~\citep{Reux2015,Boozer2015}.

 The most promising runaway-mitigation method is to inject impurities which
 dissipate the runaway beam by collisional
 scattering~\citep{HollmannDMS}. Due to the low temperatures of the
 post-disruption plasma, the impurities will only be partially
 ionized.  Since the collision frequencies 
 scale strongly
 with
 charge, the runaway dissipation rate will be heavily influenced by
 the extent to which fast electrons can penetrate the bound electron
 cloud around the impurity ion, i.e.\ the effect of \emph{partial screening}.

Partial screening has a strong effect on collision frequencies  \citep{Kirillov,mosher1975,lehtinen,dwyer,zhogolev,Hesslow}, which calls
 for accurate models of the collisional processes. Such a model
 requires a quantum-mechanical treatment of both elastic and inelastic
 collisions, as well as knowledge of the electronic charge density of
 the impurity ion. Previous treatments of partially screened elastic
 electron-ion collisions are limited to either a semi-classical
 treatment~\citep{mosher1975,martinsolis1}, or employ the Thomas--Fermi
 theory for the electron charge density~\citep{zhogolev,Kirillov}, which is limited to intermediate distances from
 the nucleus, and does not capture the shell structure of the
 ion~\citep{landauQM}. 
Therefore, in a recent paper we 
presented a 
collision operator 
based on a quantum-mechanical treatment of both elastic and
 inelastic collisions, and used density functional theory (DFT) to
 obtain the electron-density distribution of the impurity ions~\citep{Hesslow}.
This generalization of the Fokker--Planck operator to a partially ionized plasma was expressed as modifications to the deflection and slowing-down frequencies, and it was shown that both frequencies 
  increased significantly compared to the case of complete screening, already at subrelativistic energies. This generalized operator was used by \citet{Ecrit} to derive an analytical expression including the effect of screening and radiation  on the effective critical field for runaway formation and  runaway current decay.  

The present paper details the theoretical basis of the collision operator in~\citet{Hesslow} and applies it to  investigate the effects of partial screening on runaway electron dynamics.  We compare these results with the predictions from the
 approximate Thomas--Fermi theory. 
 Using the generalized collision operator, we present a detailed
 analysis of the steady-state runaway avalanche growth-rate in the
 presence of partially ionized atoms.  The increased collisional rates
 with partially ionized impurities lead to a substantially increased
 critical electric field for runaway generation
 \citep{Ecrit}. However, when the electric field is significantly
 larger than the critical field, the runaway avalanche growth rate is
 considerably higher than in the complete screening  case --
 corresponding to a fully ionized plasma with the same net ion charge.
 This behaviour, which contradicts previous predictions~\citep{Putvinski1997}, produces an additional layer
 of complexity when evaluating the effect of partially ionized
 impurities on the number of runaway electrons.

 The presence of partially ionized impurities enhances the relative
 frequency of large-angle collisions, which are beyond the
 Fokker--Planck formalism. We therefore investigate the validity of
 the Fokker--Planck operator by comparing it to the more general
 Boltzmann operator. The results show that the Fokker--Planck operator
 accurately captures the key quantities, such as the runaway density
 and current, only the synchrotron emission spectrum at large
 electric fields is slightly less accurate.  This demonstrates that the
 generalized collision operator derived here is adequate for most
 runaway studies.

The structure of the paper is as follows. Section~\ref{sec:collop}
details the derivation of the generalized collision operator for fast
electrons in the presence of partially ionized impurities.  In
section~\ref{sec:ava}, we investigate the effects of screening on the
avalanche growth rate. Section~\ref{subsec:Boltz} compares the results
obtained using the Fokker--Planck operator to the corresponding ones
using the Boltzmann operator. Finally, section~\ref{sec:concl}
summarizes our conclusions.

\section{Generalized collision operator for fast electrons in a plasma with partially ionized impurities}
\label{sec:collop}
There are two types of collisions between fast electrons and partially ionized atoms: elastic collisions, where the state of the ion remains unchanged during the collision and the incident electron is only deflected with a negligible energy transfer; and inelastic collisions, where the ion is excited or further ionized, causing the incident electron to impart a fraction of its kinetic energy to the bound electrons.
For fast electrons, both types of collisions can be treated using the
Born approximation. In the case of elastic collisions, this requires
knowledge of the electronic charge density of the impurity ion, which we obtain from DFT calculations.  
In contrast, the inelastic collisions with bound electrons primarily lead to
collisional friction; the rate of pitch-angle scattering against bound
electrons is smaller  than the rate against ions by approximately a factor of the charge number (the full nuclear charge) $Z\gg 1$. This allows us to model collisions with bound electrons with Bethe's theory for the collisional stopping
power~\citep{bethe} without the need for detailed 
differential cross sections for these processes.

In both processes, the target particle can be treated as stationary 
since we consider incident suprathermal electrons. The average momentum of the bound electrons must be below the thermal electron momentum at
a given temperature if the ionization state is roughly equilibrated
with the electron temperature. Moreover, the ion thermal speed
fulfills $v_{T \rm i} \ll v_{T \rm e}$ due to the small electron-to-ion mass ratio.
Consequently, the collision operator presented here is valid for electron speeds $v$ fulfilling
\begin{enumerate}\itemsep4pt
\item $v/c \gg Z \alpha$ (the Born approximation), with
  $\alpha\approx1/137$ the fine-structure constant. 
The Born approximation may be accurate 
even at lower energies, 
as it has been experimentally verified for incident electron energies from  \unit[1]{keV} and above for argon and neon, which are particularly relevant for fusion experiments~\citep{MottMassey}. 
\item $\gamma-1 \gg I_j/(m_{\rm e} c^2)$ (Bethe's stopping power formula), where  $\gamma$ is the Lorentz factor and $I_j/(m_{\rm e} c^2)$
  is the mean excitation energy of the ion normalized to the electron
  rest energy, which is of the order $10^{-4}$ to $10^{-3}$ for argon and
  neon, increasing with ionization degree~\citep{sauer2015}.
\item $v \gg v_{T \rm i}$ (ions at rest). 
\end{enumerate}
By matching the high energy expressions describing the effects of partial screening to the completely screened
low-energy limit, where the electron only interacts with the ion
through the net ion charge number $Z_0$, we obtain a collision operator which can be  applied at
all energies, although it is known to be correct only when the conditions
above are fulfilled.

\subsection{The Fokker--Planck operator}
\label{sec:FP}
The Fokker--Planck collision operator between species $a$ and $b$ is given by
\begin{align} 
  C^{ab}  =  -\nabla_k \left(f_a
  {\left\langle \Delta p^k \right\rangle_{ab} } \right) +
   \frac{1}{2} \nabla_k
  \nabla_l  \left(f_a {\left\langle \Delta p^k \Delta p^l \right\rangle_{ab}}
  \right),
  \label{eq:FP}
\end{align}
where the term ${\langle \Delta p^k \!  \rangle_{ab} }$ represents the
average change in the $k$th component of the momentum of the incoming
electron during a collision, while
${\langle \Delta p^k \! \Delta p^l \rangle_{ab} }$ describes the
change in the tensor $p^k p^l$. Moreover, $p = \gamma v/c$, and $\nabla_k$ refers to the
momentum-space gradient operator.  
These moments are given by
\begin{align}
{\left\langle \Delta
  p^k   \right\rangle_{ab} } &=
  \int d\mathbf{p}' f_b(\mathbf{p'})  \label{eq:deltap}
 \int \frac{d\sigma_{ab}}{d\Omega} g_{\text \o} \Delta p^k
  d\Omega,
\\
{\left\langle \Delta
  p^k  \Delta p^l \right\rangle_{ab} } &=
 \int d\mathbf{p}' f_b(\mathbf{p'}) 
 \int \frac{d\sigma_{ab}}{d\Omega} g_{\text \o} \Delta p^k
 \Delta p^l  d\Omega,  \label{eq:deltapp}
\end{align} 
where $g_{\text \o} = \sqrt{(\mathbf{v-v'})^2-(\mathbf{v\times v'})^2/c^2}$ is the M\o ller relative speed and $\nofrac{d\sigma_{ab}}{d\Omega}$ is the differential scattering cross section between species $a$ and $b$.
Here, the angular integral is taken over 
\begin{equation}
\int \dd \Omega = \int_{\theta_{\rm min}}^{\pi} \sin\theta \dd \theta  \int_0^{2 \pi} \dd \phi ,
\end{equation}
where the Coulomb logarithm, a large factor which will be described in more detail in section~\ref{sec:coulog}, enters through $\ln \Lambda = \ln (\nofrac{2}{\theta_{\rm min}})$. 
The Fokker--Planck operator can formally be seen as an expansion of
the Boltzmann operator in small momentum transfers, which is motivated
by the rapid decay of the Coulomb collision differential cross section
with momentum transfer;
$\nofrac{d\sigma_{ab}}{d\Omega}\sim \sin^{\!-4}(\theta/2)$. This
grazing collision nature of Coulomb interaction translates to a
prefactor of $\ln \Lambda$ when the collision operator is evaluated
explicitly. Consequently, the Fokker--Planck operator only retains the
terms of order $\ln \Lambda$ in equation~\eqref{eq:FP}.

When species $b$ has a Maxwellian distribution, the resulting
collision operator is parametrized by the three collision frequencies
$\nu_D^{ab}$, $\nu_S^{ab}$ and $\nu_\parallel^{ab}$, describing
deflection at constant energy (pitch-angle scattering), collisional
friction, and parallel (energy) diffusion~\citep{helander}:
\begin{equation}
    C^{ab} = {\nu_D^{ab}} \mathscr{ L}(f_a) +
    \frac{1}{p^2}\frac{\partial}{\partial p}\left[p^3 \left( {
        \nu_S^{ab}} f_a + \frac{1}{2} {\nu_\parallel^{ab} } p
      \frac{\partial f_a}{\partial p} \right)\right].
      \label{eq:collop}
\end{equation}
The pitch-angle scattering operator
$$\mathscr{L} = \frac{1}{2}\frac{\partial}{\partial
  \xi}\left(1-\xi^2\right)\frac{\partial}{\partial \xi},$$
represents scattering at constant energy, and is proportional to the
angular part of the Laplace operator. Here it is specialized to
azimuthally symmetric systems, and
$\xi = \mathbf{p}\cdot~\!\!\!\mathbf{B}/(pB)$ is the cosine of the
pitch-angle with respect to a preferred direction, set here by an
applied magnetic field $\bf B$.

 \subsection{ The Coulomb logarithm}
 \label{sec:coulog}
 The Coulomb logarithm $\ln \Lambda$ determines a minimum scattering
 angle below which Debye shielding screens out long-range
 interaction. Furthermore, it quantifies the dominance of small-angle
 collisions compared to large-angle collisions, and therefore provides
 a measure of the validity of the Fokker--Planck operator, which only
 captures small-angle collisions accurately.  For electrons,
 $\ln \Lambda$ is the logarithm of the Debye length divided by the
 de~Broglie wavelength, which depends on the electron
 energy~\citep{SolodovBetti}.  At thermal speeds, the Coulomb
 logarithm is given by~\citep{wesson}
\begin{equation}
\ln\Lambda_0 \approx 14.9-0.5 \ln n_{\rm e20}+\ln T_{\rm keV},
\label{eq:lnL0}
\end{equation} where $T_{\rm keV}$ is the temperature in $\rm keV$ and $n_{\rm e20}$ is the free-electron density in units of $\unit[10^{20}]{m^{-3}}$.  The suprathermal expressions take the following form~\citep{SolodovBetti}: 
\begin{equation}
\begin{aligned}
\ln \Lambda^{\rm ee} &= \ln\Lambda_{\rm c} + \ln\sqrt{\gamma-1},  \\
\ln \Lambda^{\rm ei} &= \ln\Lambda_{\rm c} + \ln (\sqrt{2} p)\,,
\end{aligned}
\label{eq:lnLeei}
\end{equation} 
where we introduced a Coulomb logarithm evaluated at relativistic
electron energies:
\begin{equation}
 \lnLStar =  \ln\Lambda_0  +  \frac{1}{2}\ln\frac{m_{\rm e} c^2}{T}\approx 14.6+0.5 \ln (T_{\rm eV}/n_{\rm e20}).
 \label{eq:lnLStar}
\end{equation}
Note that the temperature dependence of $\lnLStar$ is reduced compared
to $\ln\Lambda_0$, since it describes collisions between thermal
particles and relativistic electrons as opposed to collisions among
thermal electrons. Although the energy-dependence of the Coulomb
logarithm can be neglected in many scenarios, it can be significant
for relativistic electrons at post-disruption temperatures. In such
cases, the thermal Coulomb logarithm is often on the order of
$\ln\Lambda_0 \approx 10$ while
$\tfrac{1}{2}\ln (m_{\rm e} c^2/T) \approx 5$ at $T = \unit[10]{eV}$.
It is then appropriate to use $\lnLStar$ in the relativistic collision
time: $\tau_c = (4 \pi n_{\rm e} c r_0^2 \ln\Lambda_c)^{-1}$, where
$r_0$ is the classical electron radius.

An accurate treatment of the Coulomb logarithm that can be used in the
collision operator however requires a formula that is valid from
thermal to relativistic energies.  We therefore match the thermal
Coulomb logarithm~\eqref{eq:lnL0} with the suprathermal Coulomb
logarithms~\eqref{eq:lnLeei} according to
\begin{equation}
\begin{aligned} \ln \Lambda^{\rm ee} &= \ln\Lambda_0 +
\frac{1}{k}\ln\left\{1+\left[2(\gamma-1)/p^2_{T{\rm e}}\right]^{k/2} \right\},\\ \ln \Lambda^{\rm ei} &= \ln\Lambda_0 +
\frac{1}{k}\ln\left[1+(2 p/p_{T{\rm e}})^k \right],\end{aligned}
\label{eq:lnL}
\end{equation} 
 where
$p_{T{\rm e}} = \sqrt{2T/(m_{\rm e} c^2)}$ is the thermal momentum, and the parameter $k=5$ is chosen to give a smooth transition between $\ln\Lambda_0$ and $\ln\Lambda^{\rm ee (ei)}$. The precise value of $k$ does not significantly impact the resulting runaway dynamics, but a differentiable function facilitates implementation in numerical kinetic solvers.

%%%%%%%%%%%%%%%%%%%%%%%%%%%%%%%%%%%%%%%%%%%

\subsection{Elastic electron-ion collisions}
\label{sec:nuei}

In this section, we follow the recipe of~\citet{rosenbluth} and \citet{akama} to derive a generalized collision operator that takes partial screening into account by including a more general differential cross section in equation~\eqref{eq:FP}.  We model elastic electron-ion collisions quantum-mechanically in the  Born approximation. With the ions as infinitely heavy stationary target particles initially at rest, the differential scattering cross section takes the following form \citep{MottMassey}:
\begin{equation}
\frac{d\sigma_{{\rm e}j}}{d\Omega} = \frac{r_0^2}{4 p^4} \lpa \frac{\cos^2(\theta/2) p^2 + 1 }{\sin^4(\theta/2)}\rpa  \left|Z_j-F_j(q)\right|^2,
\label{eq:crossSection}
\end{equation}
where the form factor for ion species $j$ is defined as
\begin{equation}
F_j(\mathbf{q}) =  
\int \rho_{{\rm e},j}(r) {\rm e}^{-{\rm i} \mathbf{q\cdot r}/a_0}\,\dd\mathbf{r}\,.\label{eq:F(q)}
\end{equation}
Here, $\mathbf{q} = 2 \mathbf{p} \sin(\theta/2) / \alpha $, and $a_0 = \hbar/(m_e c \alpha)$ is the Bohr radius.  The high- and
low-energy behaviour of the form factor represent the limits of
complete and no screening: at low $q$, the exponential approaches
unity and thus the form factor is to lowest order given by the number
of bound electrons $N_{{\rm e},j}$, whereas at high $q$ the fast
oscillations in the exponential instead cause the form factor to
vanish. Consequently, the factor $|Z_j-F_j|^2$ varies between the net charge number squared $Z_{0j}^2$ and the atomic number squared $Z_j^2$ of ion species $j$. The ratio
between these limits is typically of order $10^2$ for weakly ionized
high-Z impurities, which motivates an accurate description of the
effect of partial screening in the intermediate region.

We define a local center of mass frame $\{ {\mathbf e}_L^i\}$ with $p^0_L$ time-like, $e^1_L = \mathbf{p}/p$ parallel to the initial momentum, while $e^2_L$ and $e^3_L$ are orthogonal to $e^1_L$. The momentum transfers can then be written in terms of the deflection angle $\theta$ as follows:
\begin{equation}
 \begin{aligned}
 \Delta p_L^0 &= 0, \\
 \Delta p_L^1 &=  p (\cos\theta-1), \\
  \Delta p_L^2 & =p \sin\theta\cos\phi, \\
   \Delta p_L^3 &=p \sin\theta\sin\phi.
 \end{aligned}
 \label{eq:deltaPs}
\end{equation}

Inserting the cross section in equation~\eqref{eq:crossSection} and $\Delta p^k$ from equation~\eqref{eq:deltaPs} into the moments in  equations~\eqref{eq:deltap}-\eqref{eq:deltapp}, we evaluate the integral over the  azimuthal angle $\phi$. There are three non-vanishing moments: $\int_0^{2 \pi} \dd \phi = 2\pi$ and $\int_0^{2 \pi} \sin^2\!\phi\dd \phi = \int_0^{2 \pi} \cos^2\!\phi\dd \phi = \pi$, respectively corresponding to $\langle \Delta p_L^1 \rangle, 
\langle \Delta p_L^1 \Delta p_L^1\rangle$ and $
\langle \Delta p_L^2 \Delta p_L^2\rangle=
\langle \Delta p_L^3 \Delta p_L^3\rangle$.  
With species $a$ denoting electrons and the target particles $b$ denoting stationary ions of species $j$, so that $f_j(\mathbf{p}) = n_j \delta(\mathbf{p})$, the moments are given by
\begin{equation}
 \begin{aligned}
\left\langle \Delta p_L^1\right\rangle_{{\rm e}j}&= - 4 \pi n_j p v \int_{1/\Lambda}^{1} 4 \frac{d\sigma_{{\rm e}j}}{d\Omega}  x^3 \dd x,\\
\left\langle \Delta p_L^1 \Delta p_L^1 \right\rangle_{{\rm e}j}& = 8 \pi n_j p^2 v \int_{1/\Lambda}^{1} 4 \frac{d\sigma_{{\rm e}j}}{d\Omega}  x^5 \dd x, \\
\left\langle \Delta p_L^2 \Delta p_L^2 \right\rangle_{{\rm e}j}& = 4 \pi n_j p^2 v \int_{1/\Lambda}^{1} 4\frac{d\sigma_{{\rm e}j}}{d\Omega}  x^3(1-x^2)\, \dd x = \left\langle \Delta p_L^3 \Delta p_L^3 \right\rangle,
\end{aligned}
\end{equation}
where $x = \sin (\theta/2)$. Inserting the differential cross section from equation~\eqref{eq:crossSection} yields
\begin{equation}
\label{eq:<Deltap>}
 \begin{aligned}
\left\langle \Delta p_L^1\right\rangle_{{\rm e}j}&=
- 4 n_j  \pi  r_0^2 \frac{v}{p^3}  \int_{1/\Lambda}^{1}  
\frac{1}{x} \left[(1-x^2) p^2 + 1 \right] 
 \left|Z_j-F_j(q)\right|^2
 \dd x, \\
\left\langle \Delta p_L^1 \Delta p_L^1 \right\rangle_{{\rm e}j}&= 
 8  n_j  \pi  r_0^2 \frac{v}{p^2}  \int_{1/\Lambda}^{1}  
x\left[ (1-x^2) p^2 + 1 \right]  
\left|Z_j-F_j(q)\right|^2
 \dd x, \\
\left\langle \Delta p_L^2 \Delta p_L^2 \right\rangle_{{\rm e}j}&= 
4 n_j  \pi  r_0^2  \frac{v}{p^2}  \int_{1/\Lambda}^{1}\frac{1-x^2}{x} %
\left[ (1-x^2) p^2 + 1 \right]  
\left|Z_j-F_j(q)\right|^2
 \dd x % 
 = \left\langle \Delta p_L^3 \Delta p_L^3 \right\rangle.
 %\label{eq:<Deltap3Deltap3>},
\end{aligned}
\end{equation}

Unlike the non-relativistic case, the relativistic Fokker--Planck operator does not capture the correct interspecies energy transfer of the corresponding Boltzmann operator. In the case considered here, of collisions with stationary heavy targets, an unphysical non-zero energy transfer occurs. This can be avoided by expanding the integrands of \eqref{eq:<Deltap>} to leading-order in the scattering angle parameter $x$, but at the same time allowing the momentum transfer $q = 2 p x / \alpha$ to be non-negligible as it contains the large factor $p/\alpha$. The resulting form of the operator is validated against the Boltzmann operator in section~\ref{subsec:Boltz}: it is shown that
with this choice the
loss rates of parallel momentum of the Fokker--Planck and Boltzmann operators are equal
 at non-relativistic energies, and differ by a term of order $1/\ln \Lambda$ in the ultra-relativistic limit.

For the moments, we thus obtain
\begin{equation}
\begin{aligned}
\left\langle \Delta p_L^1\right\rangle_{{\rm e}j}&= - 4 \pi n_j  c r_0^2 \frac{\gamma}{p^2} [Z_0^2\ln\Lambda^{\rm ei} + g_j(p)]  
, \label{eq:<Deltap1>v2}\\
\left\langle \Delta p_L^1 \Delta p_L^1 \right\rangle_{{\rm e}j}&= 0,\\
\left\langle \Delta p_L^2 \Delta p_L^2 \right\rangle_{{\rm e}j}&= 4 \pi  n_j  c r_0^2  \frac{\gamma}{p} [Z_0^2\ln\Lambda^{\rm ei} + g_j(p)] , 
\end{aligned}
\end{equation}
where
\begin{equation}
\label{eq:g(p)}
g_j(p) \equiv \int_{1/\Lambda}^1 \frac{1}{x} \left[\left|Z_j-F_j(q)\right|^2 - Z_{0,j}^2\right] \dd x.
\end{equation}

To obtain an explicit form of the collision operator in spherical coordinates $\{p, \theta, \phi\}$, where $\mathbf{p} = (p,0,0)$, we transform the expressions in equation~\eqref{eq:<Deltap1>v2} into an arbitrary coordinate system $\{ {\mathbf e}^\mu\}$ and then evaluate the collision operator using covariant notation. For details of this calculation, we refer the reader to appendix~\ref{app:covariant}. The  collision operator then becomes
\begin{align}
 C^{{\rm e}j} &=\frac{1}{p^2 \sin \theta } \partial_\mu
\left(p^2 \sin \theta  V^\mu 
\right), \label{eq:CeiSimp}
\end{align} 
where 
\begin{align}
V^\mu =
 \begin{pmatrix}
 -\left[\left\langle\Delta p_L^1 \right\rangle_{{\rm e}j}+ 
 \frac{1}{p } \left\langle \Delta p_L^2 \Delta p_L^2\right\rangle_{{\rm e}j}
 \right] f_{\rm e} \label{eq:ZERO}\\ 
  (2 p^2)^{-1} \left\langle \Delta p_L^2 \Delta p_L^2\right\rangle_{{\rm e}j}
\partial_\theta f_{\rm e}  \\
(2 p^2\sin^2 \theta)^{-1} 
  \left\langle \Delta p_L^2 \Delta p_L^2\right\rangle_{{\rm e}j}\partial_\phi f_{\rm e}
 \end{pmatrix}^\mu .
\end{align}
From the first component of equation~\eqref{eq:ZERO}, it is clear that the contributions to the 
energy loss vanish identically only if higher-order terms in the Fokker--Planck operator are neglected so that $ \left\langle\Delta p_L^1 \right\rangle_{{\rm e}j}=- 
p^{-1} \left\langle \Delta p_L^2 \Delta p_L^2\right\rangle_{{\rm e}j}$. Finally, evaluating equation~\eqref{eq:CeiSimp} for an axisymmetric plasma yields, after summation over ion species $j$,  the electron-ion collision operator
\begin{align}
C^{\rm ei} &=  \sum_j \frac{1}{p^2}\langle \Delta p_L^2 \Delta p_L^2\rangle_{{\rm e}j}\frac{1}{2}\frac{\partial}{\partial \xi}\left(1-\xi^2\right)\frac{\partial}{\partial \xi}  f_{\rm e} \\&=
\sum_j  4 \pi n_j c r_0^2  \frac{\gamma}{p^3} [Z_0^2\ln\Lambda^{\rm ei} + g_j(p)]\mathscr{L}\{ f_{\rm e}\},
\label{eq:CeiFPfinal}
\end{align} 
and we can identify the deflection frequency
\begin{equation}
\nu_D^{\rm ei} =   4  \pi c  r_0^2  \frac{\gamma}{p^3} \bigg(n_{\rm e} Z_{\rm eff}\ln\Lambda^{\rm ei} +\sum_j n_j g_j(p)\bigg),
\label{eq:nuD}
\end{equation}
where the first term is the completely screened collision frequency with the effective charge defined as $Z_{\rm eff} = \sum _j n_j Z_{0,j}^2/n_{\rm e} $.
Note that the properties of the form factor ensure that the completely screened limit is reached if either $p \rightarrow 0$, or if the ion is fully ionized so that $Z=Z_0$.

\begin{figure}
\centering
   \includegraphics[width=0.6\columnwidth]{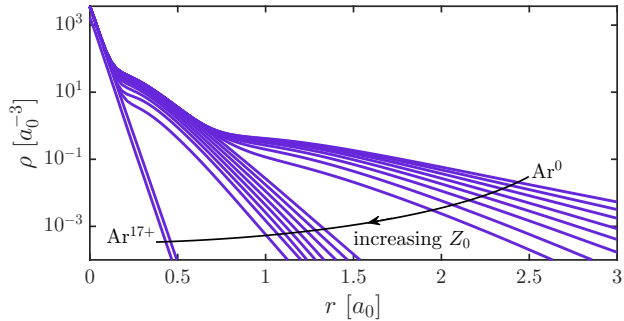}
   \caption{Number density of bound electrons averaged over solid angle as
     a function of radius for all ionization states of argon. The
     length scale is given in units of the Bohr radius $a_0$.}
    \label{fig:DFTdensity}
  \end{figure}

What remains is to find the screening function $g_j(p)$ for all ion
species $j$. This requires the electronic charge distribution of the
ion, which we determine from density functional theory
(DFT), using the programs \textsc{exciting}~\citep{excitingrevp} and
\textsc{gaussian}~\citep{g16}. The \textsc{gaussian} calculations were
performed using the hybrid-exchange correlation functional
PBE0~\citep{adamo1999}, a Douglas--Kroll--Hess second-order scalar
relativistic Hamiltonian~\citep{douglas1974,hess1986,barysz2001}, and
the atomic natural orbital-relativistic correlation consistent basis
set, ANO-RCC~\citep{widmark1990,roos2004,roos2005}.  As an example,
figure~\ref{fig:DFTdensity} shows the density of bound electrons as a
function of radius for all argon ionization states. Note that the
density decay can be approximately parametrized with piecewise
exponentials having different slopes for each of the atomic shells.

  When calculating the form factor, the electronic density was first
  spherically averaged, in which case the form factor in
  equation~\eqref{eq:F(q)} simplifies to
\begin{equation}
F_j(q) =  
4 \pi \int_0^\infty \rho_{{\rm e},j}(r) \frac{r a_0}{ q}\sin(q r/a_0)\,\dd r\,, \label{eq:F(q)2}
\end{equation}
where again $q = 2 p x / \alpha$ and the total number of bound
electrons is given by $N_{\rm e} = 4 \pi\int r^2 \rho_{{\rm e},j}(r) \dd r$.

Numerically, we find that the form factor is well described by a
generalized version of the form factor obtained from the Thomas--Fermi
model by~\citet{Kirillov}:
\begin{equation}
F_{j,\mathrm{\textsc{tf-dft}}}(q) = \frac{N_{{\rm e},j}}{1+(q a_j)^{3/2}}. \label{eq:FKr}
\end{equation}
Note that we can extend the lower integration limit to zero in
the definition of $g_j(p)$ \eqref{eq:g(p)} since the integrand is finite as
$p\rightarrow 0$ (the logarithmically diverging terms cancel as shown
in appendix~\ref{app:gHighEnergy}). In the form factor in
equation~\eqref{eq:F(q)2}, this extension of the integral amounts to
neglecting terms of order $\Lambda^{-3/2}\ll 1$ and $(p \bar a_j/ \Lambda)^{3/2} \ll 1$ which
describe the transition from partial screening to no
screening. However, since $\Lambda^{\rm ei} = \exp(\ln \Lambda^{\rm ei}) \propto p$ at high energies from equation~\eqref{eq:lnLeei}, we obtain $(p \bar a_j/ \Lambda)^{3/2} \sim 137/\Lambda_c$; therefore, this approximation is always valid and the no screening
limit will never be reached. Equation~\eqref{eq:FKr} then gives
\begin{equation}
g_j(p) = \frac{2}{3}(Z_j^2-Z_{0,j}^2)\ln[(p \bar a_j)^{3/2}+1]-\frac{2}{3} \frac{N_{{\rm e},j}^2 (p \bar a_j)^{3/2}}{(p \bar a_j)^{3/2}+1}.
\label{eq:g}
\end{equation}
This model, which we denote the Thomas-Fermi--DFT (TF-DFT) model,
includes one free parameter: the effective ion length scale $a_j$ in
units of the Bohr radius $a_0$, with $\bar a_j= 2 a_j/\alpha$. This
parameter is determined from the density of bound electrons obtained
from the DFT calculations.

\begin{figure}\centering 
\includegraphics[width=0.6\linewidth]{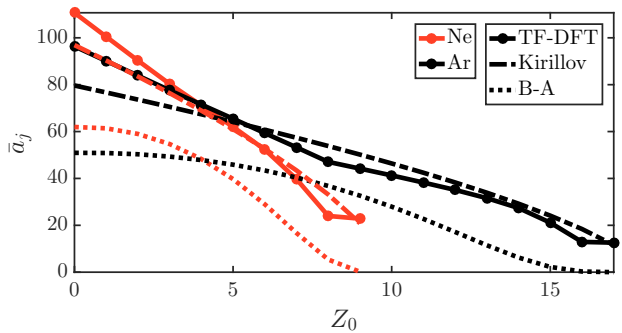}
\caption{\label{fig:a}Length-scale $a_j$ for Ne and Ar, compared to both the Thomas--Fermi model with the Kirillov solution from equation~\eqref{eq:aTF}, and the Breizman--Aleynikov (\zkba) model from equation~\eqref{eq:aBA}. Note that by  definition, $\bar a_j = \bar a_\textnormal{\sc tf-dft} \equiv \bar a_\textnormal{\sc dft}$.}
\end{figure}

The general properties of the screening function $g_j(p)$ allow us to
determine $a_j$ so that the deflection frequency exactly matches the
high-energy asymptote of the DFT results.  As shown in
appendix~\ref{app:gHighEnergy}, $g_j(p)$ always takes the form
\begin{equation}
g_j(p) = (Z_j^2 -Z_{0,j}^2)\ln (2p/\alpha) + C, \qquad 2 p/\alpha \gg 1
\label{eq:gAs}
\end{equation}
where only the constant $C$ depends on the specific ionic distribution. Since the additive constant can be absorbed into the effective length scale, the high-energy behaviour of the screening function is reduced to a one-parameter problem. This indicates that equation~\eqref{eq:g} should be well-suited as an analytic model of the screening problem, if it  approximates the transition from the low-momentum behaviour to the high-momentum behaviour. Accordingly,  we determine $a_j$ for an arbitrary charge distribution $\rho_{{\rm e},j}(r)$ by matching the $g_j(p)$ in equation~\eqref{eq:gAs} to the general high-energy asymptote of $g_j(p)$, 
\begin{equation}
g_j(p) \sim (Z_j^2-Z_{0,j}^2)\ln (p \bar a_j) -\frac{2}{3}N_{{\rm e},j}^2, \qquad p \bar{a}_j \gg 1.
\end{equation}
The resulting closed form of the effective length scale  $\bar a_j$ is given in equation~\eqref{eq:abar} in  appendix~\ref{app:gHighEnergy}, and tabulated for many of the fusion-relevant ion species in table~\ref{tab:consts}. The constants for argon and neon are illustrated in figure~\ref{fig:a} as a function of $Z_0$ in solid line. Curiously, the shell structure observed in the charge density of figure~\ref{fig:DFTdensity} can be discerned as discontinuities in $\partial \bar a_j/\partial Z_{0,j}$.

\begin{figure}\begin{center}
\includegraphics[width=0.95\linewidth]{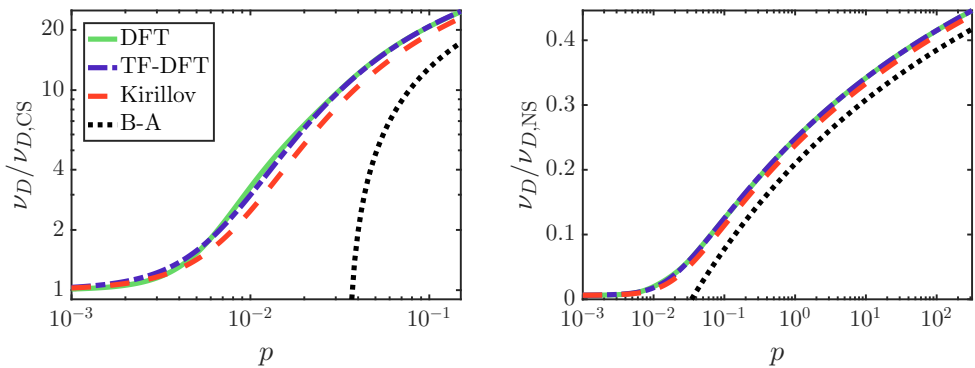}\end{center}
\caption{\label{fig:nuD} Comparison between the DFT and TF-DFT models for the   enhancement of the deflection frequency.  Left panel is shown at low energies and normalized to the completely-screened (CS), low energy limit. Right panel shows the behaviour up to higher energies, and is normalized to the no screening (NS) limit. The deflection frequency is significantly lower than the no-screening limit even at ultrarelativistic speeds. The figure is for Ar$^{1+}$, and the Coulomb logarithm was determined by setting $T=\unit[10]{eV}$ and  $n_{\rm e} = \unit[10^{20}]{m^{-3}}$.}
\end{figure}

 Since the obtained values are $\bar a_j \sim 10^2$ for several weakly ionized species such as neon and argon, the deflection frequency will be significantly enhanced compared to complete screening already at $p \sim 10^{-2}$. This is confirmed in figure~\ref{fig:nuD}, which also shows that the most accurate model for the deflection frequency -- the DFT model (solid, green line) -- is well approximated by the TF-DFT model in dash-dotted blue over the entire energy interval from non-relativistic to ultra-relativistic energies.  

The length parameter $\bar a_j$ is well suited to compare our result with previous work since it completely
characterizes the behaviour of the deflection frequency at high energy, which is the most important region for fast-electron dynamics. A comparison at low energies, where the screening function cannot in general be described by a single parameter, should be approached with caution as the Born approximation is only valid in the regime $\beta \gtrsim Z \alpha \Leftrightarrow p\gtrsim [(Z\alpha)^{-2} -1]^{-1/2} \sim 10^{-1}$. The behaviour at lower momenta is approximate, and should merely be regarded as an interpolation between the low energy limit of complete screening (which is reproduced by the TF-DFT model) and the behaviour at higher energies. Therefore, we primarily focus on the length scale $\bar a_j$ when comparing with previous work. 
 For example, the result of~\citet{Kirillov} corresponds to 
\begin{equation}
\bar a_{\rm Kirillov} =  \frac{2}{\alpha} \frac{\left(9 \pi\right)^{1/3}}{4}  \frac{N_{\rm e}^{2/3}}{Z} \approx  \frac{2}{\alpha}\frac{3}{4}\frac{N_{\rm e}^{2/3}}{Z}\label{eq:aTF}.
\end{equation}
The Kirillov model captures the approximate 
scaling of $\bar a_j$ with $Z$ and $Z_0$, however it differs significantly from the DFT results at low ionization degrees (maximum relative error 20\%, obtained for C$^0$) and for $N_{\rm e}  = 2$ (maximum 43\%, Ar$^{16+}$). 
As shown in figure~\ref{fig:a}, this is because the Kirillov model does not capture the shell structure of the ion, which is an inherent characteristic of the Thomas--Fermi theory employed by \citet{Kirillov}. 
Although these relative errors are significant, the final error in the deflection frequency is modest at high energies, since the deflection frequency is only sensitive to $\ln \bar{a}_j$. At $p=0.1$, the relative error of $\bar{a}_j$ between the TF-DFT model and the Thomas-Fermi model is at most 14\%.

We find a significantly larger difference between our model for the
deflection frequency and the model used by~\citet{BreizmanAleynikov2017Review}. 
In this model, which we refer to as the
\zkba\ model, the deflection frequency always increases
logarithmically. The deflection frequency therefore diverges as $p\rightarrow 0$ and
the complete screening limit is consequently not reproduced, which is
illustrated in figure~\ref{fig:nuD}a.  This means that the \zkba\
model is only applicable at relativistic energies and is unable to
describe phenomena involving mildly relativistic electrons, such as
hot-tail, primary runaway generation and the avalanche
mechanism at high electric fields. In the B--A model, the logarithmic increase of the
deflection frequency corresponds to the length constant
\begin{align}
\bar a_{\textsc{\zkba}} = \frac{2}{\alpha}Z_j^{-1/3} 
\exp\left( \frac{2}{3}\frac{N_{{\rm e},j}^2-6\ln 2(Z_j Z_{0,j} -Z_j^2-Z_{0,j}^2)}{Z_j^2-Z_{0,j}^2}\right).\label{eq:aBA}
\end{align}
As shown in figure~\ref{fig:a}, $\bar a_\textrm{\zkba}\!$ differs significantly from both $\bar a_{\rm Kirillov}$ and our more accurate DFT-based values of $\bar a_j$. 

We conclude that the Kirillov formula suffices for an accurate description of screening in most situations, although the constants derived from DFT have a higher level of accuracy, especially at low momenta. 

\subsection{Inelastic collisions with bound electrons}

Unlike for elastic collisions with partially screened nuclei, there is no analytic expression for the differential cross section for inelastic collisions between fast and bound electrons, but the energy loss is described by the Bethe stopping-power formula~\citep{bethe,jackson}. Accordingly, we modify the slowing-down frequency $\nu_S^{\rm ee}$ in equation~\eqref{eq:collop}, which describes collisional drag, whereas we neglect the modification of the electron-electron deflection frequency $\nu_D^{\rm ee}$, since it does not follow from the stopping-power calculation. The error introduced 
through this approximation, i.e.\ $\nu_D \approx \nu_{D}^{\rm ei}+ \nu_{D, \textsc{cs}}^{\rm ee}$, 
can be estimated by considering the limits of no screening and complete screening of $\nu_D^{\rm ee}$. For suprathermal electrons, $\nu_{D, \textsc{cs}}^{\rm ee} =   4  \pi c  r_0^2  (\nofrac{\gamma}{p^3}) n_{\rm e} \ln\Lambda^{\rm ee}$, while $\nu_{D, \textsc{ns}}^{\rm ee}$ is enhanced by a factor  of 
$n_{\rm e}^{\rm tot}/n_{\rm e} = 1 + \sum_j N_{{\rm e},j} n_j/n_{\rm e}$.
Comparing to the electron-ion deflection frequency~\eqref{eq:nuD}, we find that our approximation is valid if either $\sum_j Z_{j}^2 n_j \gg \sum_j N_{{\rm e},j} n_j$, or if $ 1+Z_{\rm eff} \gg \nu_{D}^{\rm ee}/\nu_{D, \textsc{cs}}^{\rm ee}$ due to either significant ionization levels or low electron momentum. In other words, our model is accurate both when screening effects are small and in the presence of high-Z impurities.

\begin{figure}\begin{center}
\includegraphics[width=0.6\linewidth]{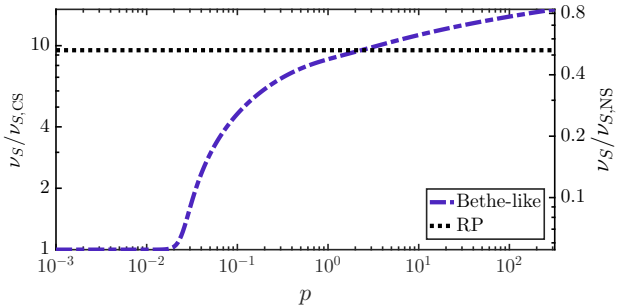}
\end{center}
\caption{\label{fig:nuS} The partially screened slowing-down frequency for the Bethe-like model in equation~\eqref{eq:nuS} and the RP model from equation~\eqref{eq:nuSRP}, for singly ionized argon. The collision frequency is normalized to the completely screened (CS), low-energy limit on the left y-axis, and to the limit of no screening (NS) on the right y-axis. The figure is for Ar$^{1+}$, and the Coulomb logarithm was determined by setting $T=\unit[10]{eV}$ and  $n_{\rm e} = \unit[10^{20}]{m^{-3}}$.
}
\end{figure}

The Bethe stopping-power formula modifies the slowing-down frequency $\nu_S^{\rm ee}$ describing collisional drag according to~\citet{bethe} and \citet{jackson}
\begin{equation}
 \nu_S^{\rm ee} = 
  4  \pi c  r_0^2  \frac{\gamma^2}{p^3}
\bigg[ n_{\rm e} \ln\Lambda^{\rm ee} +
\sum_j n_j  N_{{\rm e},j} 
\left( \ln h_j-\beta^2\right)\bigg],
\label{eq:bethe}
\end{equation}
where
 $h_j = p\sqrt{\gamma-1}(m_{\rm e} c^2 /I_j)$,  and $I_j$ is the mean excitation energy of the ion.
In this work, the numerical values of $I_j$ for different ion species were obtained from \citet{sauer2015}. In addition, several sources list the mean excitation energy for neutral atoms, for instance \citet{ICRU}, which is used in \textsc{estar}~\citep{ESTAR}.
Equation~\eqref{eq:bethe} is valid for $m_{\rm e} c^2(\gamma-1) \gg I_j$, which is typically on the order of hundreds to thousands of eV. In order to find an expression that is applicable over the entire energy range from thermal to ultrarelativistic energies, we match equation~\eqref{eq:bethe} to the low-energy asymptote corresponding to complete screening. The resulting interpolation formula, which we refer to as the \emph{Bethe-like model}, is given by
\begin{align}
 \nu_S^{\rm ee} =& 
  4  \pi c  r_0^2  \frac{\gamma^2}{p^3}
\bigg\{ n_{\rm e} \ln\Lambda^{\rm ee} +
\sum_j n_j  N_{{\rm e},j} 
\bigg[ \frac{1}{k}
\ln\left( 1+ h_j^k\right)-\beta^2\bigg]\bigg\},
\label{eq:nuS}
\end{align}
where we set $k=5$. This is plotted as a function of momentum in figure~\ref{fig:nuS}, and compared to the completely screened  limit on the left y-axis, and the limit of no screening  on the right y-axis. 
Unlike the deflection frequency, equation~\eqref{eq:nuS} will exceed the limit of no screening in the limit of infinite momentum, since it increases by a power of $p^{3/2}$ compared to a power of $p^{1/2}$ for $\ln\Lambda^{\rm ee}$ in equation~\eqref{eq:lnLeei}. For fusion-like densities, this will however happen around $p\sim 10^{4}$ ($\unit[\sim\! 10]{GeV}$), which is well above realistic runaway energies.
At these ultra-large momentum scales, the so-called density effect~\citep{SolodovBetti,jackson} would ensure that the logarithmic term smoothly approaches the Coulomb logarithm.

We also compare the Bethe-like model to the Rosenbluth--Putvinski (RP) model~\citep{RosenbluthPutvinski1997}, which includes half of the bound electron density $n_{\rm b} = \sum_j n_j  N_{{\rm e},j}$:
\begin{equation}
 \nu_S^{\rm ee} \approx  
  4  \pi c  r_0^2  \frac{\gamma^2}{p^3}\ln\Lambda
\left( n_{\rm e}  +\frac{n_{\rm b}}{2} \right).
\label{eq:nuSRP}
\end{equation}
Figure~\ref{fig:nuS} shows that this estimate coincides with the Bethe-like model at $p\approx 1$, but results in a notable overestimation at mildly relativistic momenta and a significant underestimation at ultra-relativistic momenta.

Note that equation~\eqref{eq:nuS}  ensures that the enhancement of $\nu_S^{\rm ee}$ does not extend into the bulk electron population, which means that the 
first term $4  \pi c  r_0^2  (\nofrac{\gamma^2}{p^3}) n_{\rm e} \ln\Lambda^{\rm ee}$ 
can be replaced by the
complete expression for $\nu_{S,\textsc{cs}}^{\rm ee}$ accounting for a finite bulk temperature~\citep{BraamsKarney}. 
 This is because $I_j$ is greater than the temperature $T$ at which a certain ion species $j$ would be present in equilibrium. 
  Since the ions can always be treated as stationary (at rest), the same issue does not arise for $\nu_D^{\rm ei}$. This means that 
the generalization of the Fokker--Planck operator to a partially ionized
plasma can be expressed as modifications to $\nu_D^{\rm ei}$ and $\nu_S^{\rm ee}$
in the collision operator~\eqref{eq:collop}, 
  according to equation~\eqref{eq:nuD}, with $g_j(p)$ defined in equation~\eqref{eq:g} and $\bar a_j$ given in table~\ref{tab:consts}, as well as equation~\eqref{eq:nuS}, with $I_j$ from \citet{sauer2015}.

\section{Effect on avalanche growth rate and runaway distribution}
\label{sec:ava}
The presence of partially ionized atoms has a peculiar effect on the avalanche growth rate at high electric fields: as will be shown in the present section, the partial-screening effect can increase the avalanche growth rate despite the increased collisional damping and in contrast to previous predictions~\citep{Putvinski1997}. Moreover, the quasi-steady-state runaway distribution acquires an electric field-dependent average energy since the growth rate no longer depends linearly on the electric field.  

\begin{figure}\centering
    \includegraphics[width=0.6\columnwidth]{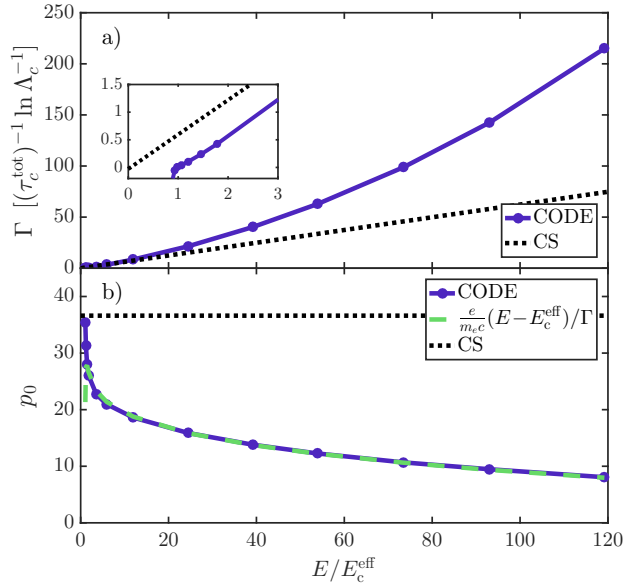}
    \caption{ a) Steady-state runaway growth rate as a function of normalized electric field. The partially screened growth rate (solid line) exceeds  the completely screened limit (dotted line) at high electric fields, but is significantly lower in the near-critical electric-field region, which is shown in the insert. 
 b) With partial screening (solid line), the average momentum $p_0$ decreases with electric field, as predicted by the green  dashed line,  and is lower than in the completely screened limit (dotted line).  The simulation was done at $T=\unit[10]{eV}$ with a plasma composition of D and Ar$^{1+}$, where $n_{\rm D} =  \unit[10^{20}]{m^{-3}}$ and  $n_{\rm Ar} = 4 n_{\rm D} $.}
    \label{fig:ava}
  \end{figure}

The avalanche growth rate is defined as
\begin{equation}
\Gamma  =\frac{1}{n_{\rm RE}}\frac{\dd n_{\rm RE}}{\dd t}. 
\label{eq:Gamma}
\end{equation}
With constant background parameters, the runaway distribution reaches a quasi-steady state and the avalanche growth rate approaches a constant value. This quasi-steady-state growth rate is shown in the presence of singly ionized argon impurities in figure~\ref{fig:ava}a. Here, the growth rate is plotted against $E/\Eceffeq$, where the effective critical electric field $\Eceffeq \gtrsim \Ectot = E_{\rm c} \nofrac{n_{\rm e}^{\rm tot}}{n_{\rm e}} $ is given in \citet{Ecrit}. These results were obtained by solving the
  kinetic equation using the numerical solver \textsc{code}~\citep{CODE,Stahl2016}, including avalanche generation   using the field-particle Boltzmann operator given in equation~(2.17) of \citep{olaknockon2018}, which was also studied by~\citet{chiu}. Since we here focus on electric fields well above the critical electric field, which are associated with low critical momenta, synchrotron and bremsstrahlung 
  radiation losses are neglected as they are important only at highly relativistic energies;  \citet{Ecrit} demonstrated that radiation losses only have an appreciable effect near the effective critical electric field.   
The
  parameters are characteristic of a post-disruption tokamak plasma:
   temperature $T=\unit[10]{eV}$, and density of singly ionized argon  $n_{\rm Ar}=4 n_{\rm D}$ with $n_{\rm D} = \unit[10^{20}]{m^{-3}}$. 

As shown in figure~\ref{fig:ava}a, the partially screened avalanche growth rate is non-linear in the electric field. We attribute this non-linearity to the energy-dependent enhancement of the collision frequencies. At weak electric fields, the critical momentum is large, and therefore also the enhancement of the collision frequencies; however, at larger electric fields, the critical momentum is reduced and the collision frequencies approach the completely screened value. This leads to an avalanche growth which increases faster than $\Gamma \propto E-\Eceffeq$. 

Interestingly, this non-linearity of the growth rate causes the partially-screened avalanche growth rate to \emph{exceed} the completely-screened limit at large electric fields. For the completely-screened limit, we use the Rosenbluth--Putvinski growth-rate formula~\citep{RosenbluthPutvinski1997}, which has been shown to be accurate to around \unit[10]{\%} in the fully ionized case~\citep{olaknockon2018} and is given by 
\begin{align}
\Gamma_\textsc{rp,cs} &= 
\frac{1}{\tau_c\ln \Lambda_c}\sqrt{\frac{\pi}{3( Z_{\rm eff} + 5)}}
\left(\frac{E}{E_c}-1\right)\left( 1- \frac{E_{\rm c}}{E}+ \frac{4 \pi (Z_{\rm eff}+1)^2}{3(Z_{\rm eff}+5)(E^2/E_{\rm c}^2 +3)}\right)^{-1/2}
\label{eq:GammaRPCSFULL} \\
&\approx 
\frac{1}{\tau_c\ln \Lambda_c}\sqrt{\frac{\pi}{3( Z_{\rm eff} + 5)}}
\left(\frac{E}{E_c}-1\right)
\,,\qquad E/E_{\rm c} \gg  2\!\sqrt{Z_{\rm eff}+1} .
\label{eq:GammaRPCS}
\end{align}
In figure~\ref{fig:ava}a, it is shown that the partially ionized growth rate is considerably higher than the completely screened value at large electric fields, even though it is significantly lower close to the critical electric field which is illustrated in the zoomed insert.

The enhancement of the avalanche growth rate in the presence of partially ionized atoms originates from the increased number of possible runaway electrons: since the binding energy is negligible compared to the critical runaway energy, the free and the bound electrons have equal probability of becoming runaways through close collisions. At high electric fields, this large enhancement by a factor of $n_{\rm e}^{\rm tot}/n_{\rm e}$ dominates over the increased rate of collisional losses, which sets the threshold energy for an electron to become a runaway.

The fact that partially screened impurities can lead to a reduction of the  avalanche growth at low electric fields, but an enhancement at larger electric fields, is not captured by the partially-screened Rosenbluth--Putvinski formula \citep{RosenbluthPutvinski1997,Putvinski1997}
\begin{align} 
\Gamma_\textsc{rp} = 
\frac{1}{\tau_c\ln \Lambda_c}\frac{n_{\rm e}^{\rm tot}}{n_{\rm e}}
\sqrt{\frac{\pi}{3( Z_{\rm eff}^\textsc{rp} + 5)}}\left(\frac{E}{E_{\rm c}^\textsc{rp}}-1\right),
\label{eq:GammaRP}
\end{align}
where the effective field includes half of the bound electron density $n_{\rm b}$,  originating from the same factor in $\nu_S^{\rm ee}$ from equation~\eqref{eq:nuSRP}:  $$ E_{\rm c}^\textsc{rp} =  \left(1+\frac{ n_{\rm b}}{2 n_{\rm e}}\right) E_{\rm c}, $$ and the partially ionized effective charge $Z_{\rm eff}^\textsc{rp}$ is taken from Parks--Rosenbluth--Putvinski~\citep{Parks1999}:
\begin{equation}
Z_{\rm eff}^\textsc{rp} = \sum_{\substack{j\, \rm part. \\ \rm ionized }} \frac{n_j}{n_{\rm e}} \frac{Z_j^2}{2} + \sum_{\substack{j\, \rm fully \\ \rm ionized }} \frac{n_j}{n_{\rm e}}Z_j^2.
\end{equation}
For large electric fields, $E\gg E_{\rm c}^\textsc{rp}$, and if the plasma is dominated by a weakly ionized, high-Z impurity such as Ar$^{1+}$, one obtains
\begin{equation}
\frac{\Gamma_\textsc{rp}}{\Gamma_\textsc{rp,cs}} \approx \frac{n_{\rm e} + n_{\rm b}}{n_{\rm e} + \frac{1}{2}n_{\rm b}}\sqrt{\frac{Z_{\rm eff}+5}{Z_{\rm eff}^\textsc{rp}+5}}<1.
\end{equation} 
In this case, partially ionized impurities decrease the avalanche growth rate significantly, although we find the opposite behaviour with our more accurate kinetic model: 
\begin{equation}
\Gamma > \Gamma_\textsc{rp,cs} > \Gamma_\textsc{rp}, \qquad E\gg \Eceffeq.
\end{equation}
Finally, we note that the avalanche growth rate in figure~\ref{fig:ava}a may be approximated by a second-order polynomial. This behaviour is somewhat similar to the quadratic behavior of the full Rosenbluth--Putvinski formula~\eqref{eq:GammaRPCSFULL} in the limit $2\!\sqrt{Z_{\rm eff}+1} \gg E/E_{\rm c} \gg 1$. 
However, evaluating this criterion with $Z_{\rm eff}^\textsc{rp}$ and $E_{\rm c}^\textsc{rp}$ predicts that this quadratic regime should only occur if  $E \lesssim  
9 \Eceffeq$ for the range of parameters in figure~\ref{fig:ava}.
Consequently, the Rosenbluth--Putvinski formula cannot easily be modified to accurately capture the effect of screening on the avalanche growth rate.

The increased growth rate has direct implications for the avalanche multiplication factor, which determines the maximum amplification of a small seed due to avalanche multiplication. To estimate this effect we consider the example of a tokamak disruption, where a part of the initial current  is converted to runaways via avalanching.  We follow the calculation of~\citet{helander2002} under the approximation $\Gamma \approx \Gamma_0 E/\Eceffeq$ where $\Gamma_0$ is independent of the electric field. 
Neglecting electric-field diffusion~--~which may however significantly affect the final runaway current profile~\citep{Eriksson2004,Smith2006GO}~--~the zero-dimensional induction equation is
$$  E = - \frac{L}{2 \pi R}\frac{\dd I}{\dd t},$$
where $L\!\sim\!\mu_0 R$ is the self-inductance and $R$ is the major radius
of the tokamak. Then,   
 equation~\eqref{eq:Gamma} can be written
$$ \frac{\dd}{\dd t}\ln n_{\rm RE} \approx - \frac{\dd}{\dd t} \frac{  I L \Gamma_0}{2 \pi R \Eceffeq},$$
and therefore an initial seed $n_{0}$ can be multiplied by up to a factor of 
$$  \frac{n_{\rm RE}}{n_0} = \exp \left(\frac{    I_0 L \Gamma_0}{2 \pi R \Eceffeq}  \right).$$
The exponent can be large in high-current devices~\citep{RosenbluthPutvinski1997}. 
Consequently, if the induced electric field is much larger than \Eceff, heavy-impurity injection can increase the avalanche multiplication factor significantly.
However, to fully understand runaway beam formation in the presence of partially ionized impurities, the combined effect of avalanche multiplication and seed generation  must be accounted for, as the seed formation is also sensitive to the injected impurities~\citep{Aleynikov2017}. 

 The non-linear avalanche growth rate also manifests itself in the quasi-steady-state avalanche distribution, which can be seen by following the derivation of the avalanching distribution in the limit $E \gg E_c$ by~\citet{fulop2006}, which we detail in appendix~\ref{app:spectrum}. Analogously to \citet{fulop2006}, the resulting energy-dependence of the distribution function $F(p, t) \approx 2  \pi p^2 \int_{-1}^1  f \dd \xi$ is given by 
  \begin{equation}
  F(p,t) = n_{\rm RE}(t) \frac{1}{p_0}{\rm e}^{-p/p_0},
  \label{eq:F}
  \end{equation}
where the average momentum is given by  
 $$p_0 =   \frac{e}{m_{\rm e} c}\frac{  E-\Eceffeq}{ \Gamma(E) }  . $$ 
 In contrast to the fully ionized result $p_0 = \sqrt{Z+5}\ln\Lambda_c $, the average momentum acquires a significant electric-field dependence in the presence of partially screened ions. This momentum dependence is shown in figure~\ref{fig:ava}b, where we find $p_0$ from fitting the high-energy part of the electron distribution to an exponential decay. This average energy obtained in the \textsc{code} simulation  agrees well with the prediction in equation~\eqref{eq:F} in the region where it is valid, i.e. $E\gg \Eceffeq$.  Note that the average energy is well below the complete screening limit shown in dotted line, where $p_0 \approx \sqrt{6}\ln\Lambda_c$. 

\section{Effect of partial screening on the validity of the Fokker--Planck operator}\label{subsec:Boltz}

\begin{figure} \centering
 \includegraphics[width=0.6\linewidth]{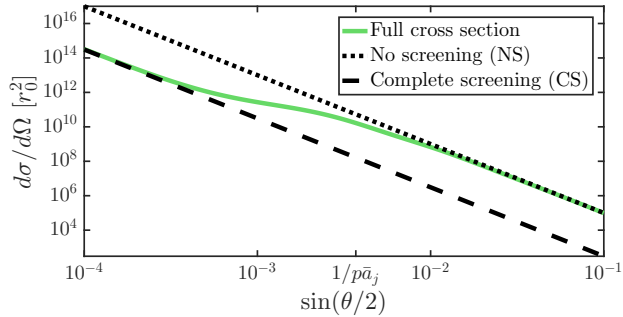}
\caption{\label{fig:crossSectionLimits} The differential
  cross section for elastic electron-ion collisions as a function of deflection angle  using the full DFT density to calculate the form factor (solid green), which exhibits a smooth transition  from complete screening (dashed black line) to the larger cross section
  with no screening (dotted black line). The cross section falls
  off as $\sin^4 (\theta/2)$;  however the curve is flatter in the transition
  region around $ \sin(\theta/2)p \bar{a}_j\sim 1$.  
  The cross section was evaluated for singly ionized argon at $p=3$.}
 \end{figure}

Scenarios where small-angle collisions dominate can be accurately
modelled by the Fokker--Planck collision operator, whereas the more
complicated Boltzmann operator must be used if large-angle
collisions are significant. Partial screening
enhances the elastic electron-ion scattering cross section for large momentum transfers while
leaving it unaltered for small momentum transfers (see
figure~\ref{fig:crossSectionLimits}). Thus, 
large-angle collisions are expected to be relatively more important in the partially screened collision
operator than in the limit of complete screening. 
In this section we will show that even though the two collision operators produce slightly different distribution functions, this difference has a negligible effect on the key runaway quantities, such as the runaway density and current.

Here, we consider the full Boltzmann operator for collisions between
runaway electrons and the background plasma. For electron-ion
collisions, we use the full operator, whereas for electron-electron
collisions, we follow the method developed by~\citet{olaknockon2018}
and only consider collisions with a momentum transfer larger than a
cutoff $p_{\rm m}$. Note that in modelling collisions with the bound
electrons, for which the full differential cross section is unknown,
the M\o ller cross section can still be used since the energy transfer
corresponding to the cutoff is typically chosen to be significantly
larger than the binding energy.

The general form of the Boltzmann operator is~\citep{cercignani}
\begin{equation}
C^{{\rm B},ab} = \int \dd \mathbf{p'} \dd \sigma_{ab}g_{\text \o} \left[ f_a(\mathbf{p}_1)f_b(\mathbf{p}_2)-f_a(\mathbf{p})f_b(\mathbf{p}') \right],
\end{equation}
where $g_{\text \o} = \sqrt{(\mathbf{v-v'})^2-(\mathbf{v\times v'})^2/c^2}$ is the M\o ller relative speed and 
$\dd \sigma_{ab}$ is the differential cross section for collisions in which the momentum of species $a$ changes from  $\mathbf{p}$ to $\mathbf{p}_1$,  while $ \mathbf{p}'\rightarrow  \mathbf{p}_2$ for species $b$.
The collision operator can be understood as the rate at which species $a$ scatters from $\mathbf{p}_1$ into $\mathbf{p}$, minus the rate of the opposite scattering process. 
Elastic electron-ion collisions are particularly convenient to model with the Boltzmann operator, since the ions can be modelled as stationary, infinitely heavy target particles and the cross section only depends on $p$, $p_1$ and $\theta$. When expanded in Legendre polynomials,
\begin{align}
C^{{\rm B},\rm ei}_{} &= \sum_j \sum_L C_L^{{\rm B,e}j} P_L(\xi)\\
f_{\rm e}(p, \theta, t) &= \sum_L f_L(p, t) P_L(\xi),
\end{align}
the Boltzmann operator takes the following form:
\begin{align}
C_L^{{\rm B,e}j} &= - n_j v f_L \int_{\theta_{\rm min}}^{\pi}  \left[1-P_L(\cos \theta) \right]\frac{\partial \sigma_{{\rm e}j}}{\partial \Omega}\dd \Omega
\\
&=
 - 2\pi n_j  c  r_0^2f_L\frac{\gamma }{   p^3} \int_{1/\Lambda}^{1} \frac{\left|Z_j-F_j(q)\right|^2}{x} 
  \frac{1-P_L(1-2x^2)}{x^2}
 \frac{(1-x^2) p^2 + 1 }{p^2+1}  
 \dd x,
\end{align}
where we again introduced $x = \sin(\theta/2)$ and inserted the differential cross section in equation~\eqref{eq:crossSection}. Using $\mathscr{L}\{ f_{\rm e}\}= -\frac{1}{2}\sum_L L(L+1)P_L(\xi)f_L$, we arrive at the following ratio between the Boltzmann operator and the Fokker--Planck electron-ion collision operator in equation~\eqref{eq:CeiFPfinal}:
\begin{align}
\frac{C_L^{{\rm B,e}j}}{C_L^{{\rm FP,e}j}} &=
  \bigg(
  \int_{1/\Lambda}^1 \frac{[Z_j-F_j(q)]^2}{x} \dd x 
 \bigg)^{-1}\!
\int_{1/\Lambda}^1 
\frac{[Z_j-F_j(q)]^2}{x} 
\frac{1-P_L(1-2 x^2)}{L(L+1) x^2}
\frac{(1-x^2)p^2+1}{p^2+1}
 \dd x \, 
 .
 \label{eq:BoltzvsFP}
\end{align}
Since $P_1(x) = x$, equation~\eqref{eq:BoltzvsFP} evaluates to unity for $L=1$ and $p=0$. Note that the same is true for the integrand when $x\ll 1$ $\forall L, p$. 

   \begin{figure}\centering
    \includegraphics[width=0.6\columnwidth]{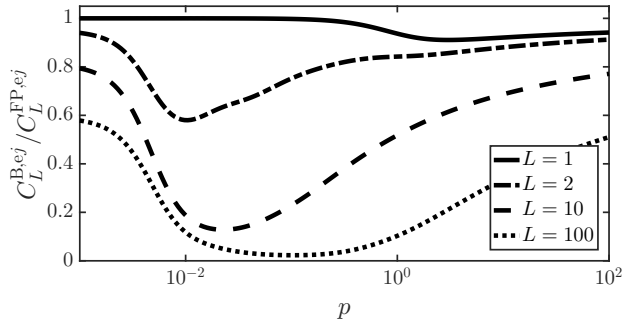}
    \caption{Ratio of the Legendre-modes of the Boltzmann and
      Fokker--Planck operators for singly ionized argon. The full DFT model was used in the figure, but the results are similar if the TF-DFT model is used instead.}
    \label{fig:BoltzvsFP}
  \end{figure}

Like the Fokker--Planck operator, the Boltzmann operator drives the distribution towards spherical symmetry, which can be seen by noting that $C_L^{{\rm B,e}j}$ is negative and proportional to $f_L$, while $C_0^{{\rm B,e}j}=0$. 
Effectively, the Boltzmann operator takes the form of a generalized $\nu_D^{\rm ei}$ which depends on the Legendre mode number $L$. The ratios
of the Legendre modes of the Boltzmann and Fokker--Planck operators are
shown in figure~\ref{fig:BoltzvsFP} for four different values of  $L$. 
As expected from equation~\eqref{eq:BoltzvsFP}, the Boltzmann operator produces the same result as the Fokker--Planck operator for $L=1$ and $p \ll 1$, and only differs by a factor of order $1/\ln\Lambda$ at higher energies. In contrast, the ratio between the Boltzmann operator and the Fokker--Planck operator decreases rapidly with $L$, and the diffusion rates are significantly reduced for $L \geq 10$ for a large range of momenta. High-$L$-structure will therefore be suppressed too quickly by the Fokker--Planck operator compared to the more accurate Boltzmann operator. This means that the two operators can be expected to produce different pitch-angle distributions in scenarios where the average pitch angle  is small.

\begin{figure}\centering
    \includegraphics[width=0.6\columnwidth]{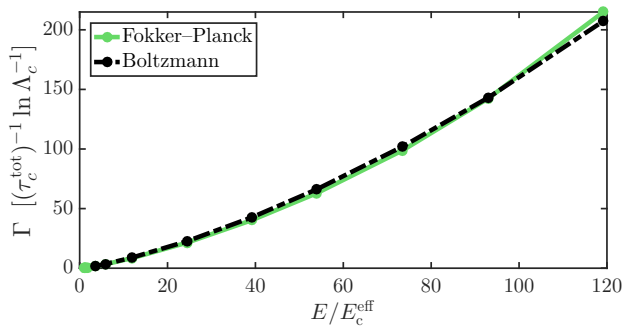}
    \caption{Steady-state avalanche growth rate as a function of normalized electric field. The Fokker--Planck and Boltzmann operators give almost identical results. The simulation was done at $T=\unit[10]{eV}$ with a plasma composition of D and Ar$^{1+}$, where $n_{\rm D} =  \unit[10^{20}]{m^{-3}}$ and  $n_{\rm Ar} = 4 n_{\rm D} $. }
    \label{fig:b2}
  \end{figure}

A suitable scenario to study the effect of the Boltzmann operator is the avalanche growth rate at high electric fields, which gives a narrow distribution function and thus requires a large number of Legendre modes to describe the distribution. 
Figure~\ref{fig:b2} shows the steady-state runaway growth
  rate as a function of $E/E_{\rm c}^{\rm eff}$ where
  $E_{\rm c}^{\rm eff}$ is the effective critical field given by \citet{Ecrit}. These growth rates were obtained by solving the   kinetic equation using \textsc{code} with the same parameters as in figure~\ref{fig:ava}, with both the Fokker--Planck operator and
  the Boltzmann operator. 
 As we show in figure~\ref{fig:b2}, the difference in the runaway growth rate between the Fokker--Planck operator and the Boltzmann operator is relatively small. 
 This result may appear surprising, since the avalanche growth rate formula~\eqref{eq:GammaRPCS} depends on $Z$, indicating a sensitivity to the pitch-angle dynamics. We speculate that the similarity  can be attributed to the agreement in the  zeroth and first Legendre modes  of the Fokker--Planck  and  Boltzmann operators as shown in figure~\ref{fig:BoltzvsFP}.  This may be sufficient since the essential runaway quantities are most sensitive to the behaviour of these modes, with the runaway density and energy fully contained in $f_0$, and the current in $f_1$.

   \begin{figure}\centering
    \includegraphics[width=0.6\columnwidth]{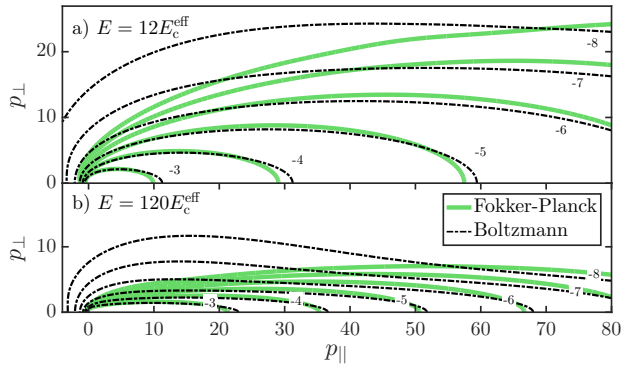}
    \caption{Contour plots of the quasi-steady-state runaway electron distribution function obtained using the Fokker--Planck operator (solid green) and the Boltzmann operator (dash-dotted, thin black), respectively. The contours show $\log_{10}(F) = (-8,-7, \dots, -3)$ as indicated in the figure, where $F =m_{\rm e}^3 c^3 f_{\rm e}/n_{\rm RE}$, so that $\int 2\pi p_\perp F \dd p_\perp \dd p_\parallel  = 1$ when integrated over the runaway population. The distributions are taken from the data points (a) $E = 12 \Eceffeq$ and (b) $E = 120 \Eceffeq$ in figure~\ref{fig:b2}.}
    \label{fig:b3}
  \end{figure}

\begin{figure}\centering
    \includegraphics[width=0.6\columnwidth]{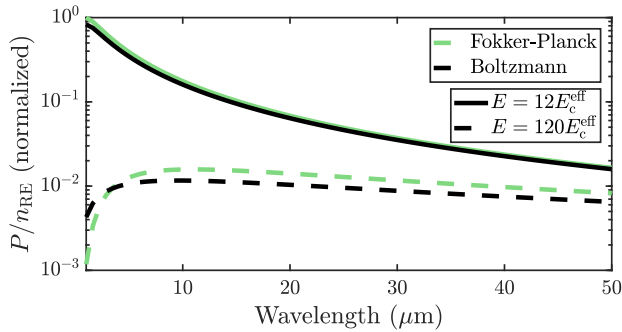}
    \caption{Synchrotron radiation spectra from the runaway electron distribution function, comparing the Boltzmann collision operator with the Fokker--Planck collision operator, in a magnetic field with strength $B=\unit[5]{T}$. Both are normalized to the maximum value of the Fokker--Planck spectrum in the chosen wave length interval.  As in figure~\ref{fig:b3}, the distributions are taken from (a) $E = 12 \Eceffeq$ and (b) $E = 120 \Eceffeq$ in figure~\ref{fig:b2}. The Boltzmann collision operator causes significantly stronger synchrotron emission than the Fokker--Planck operator, although the shape of the spectra are similar. \label{fig:b4}} 
  \end{figure}

  Figure~\ref{fig:b3} shows contour plots of the runaway electron
  distribution function using the Fokker--Planck and Boltzmann
  operators respectively. While the overall shape and energy of the distributions are similar, the Boltzmann operator leads to a pitch-angle distribution which 
develops ``wings'' consisting of a small runaway population with significantly enhanced perpendicular momentum. This effect is particularly  pronounced at high electric fields where the average pitch angle is small and at moderate energies, which is consistent with our expectation based on figure~\ref{fig:BoltzvsFP}.
  This
 indicates that using the Boltzmann operator could affect
  quantities that are particularly sensitive to the angular distribution,
  such as the emitted synchrotron radiation~\citep{Finken1990,SOFT1,SOFT2}. In order to
  quantify the differences we used the \textsc{syrup} code~\citep{Stahl2013} to calculate synchrotron spectra from the
  runaway electron distributions using the Fokker--Planck and Boltzmann
  operators, respectively, with a $\unit[5]{T}$ magnetic field.
  Figure~\ref{fig:b4} shows that in comparison with the Fokker--Planck operator, the Boltzmann collision operator leads to a spectrum with peak at a shorter wavelength. Again, we see that the difference is more pronounced at larger electric fields.
  
\begin{figure}\centering
    \includegraphics[width=0.6\columnwidth]{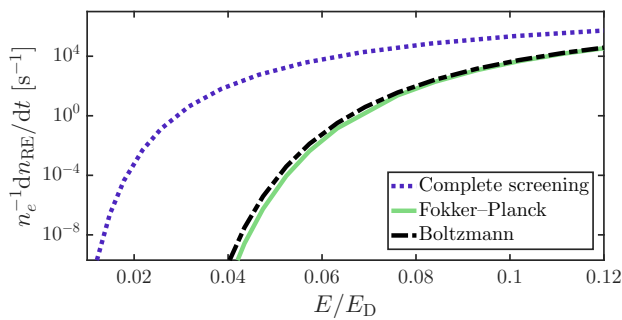}
    \caption{Steady-state primary growth rate as a function of the electric field normalized to the Dreicer field (calculated with the free electron density). 
Screening effects lead to significantly lower growth rates than the completely screened dotted blue line, but the Fokker--Planck operator (solid green) and Boltzmann operator (dash-dotted black) show a qualitatively similar behaviour.   
 The simulation was done at $T=\unit[10]{eV}$ with a plasma composed of D and Ar$^{1+}$, where $n_{\rm D} =  \unit[10^{20}]{m^{-3}}$ and  $n_{\rm Ar} = 4 n_{\rm D} $. }
    \label{fig:Dreicer}
  \end{figure}

Another quantity which is highly sensitive to input parameters is the primary (Dreicer) growth rate, which in a fully ionized plasma varies exponentially with both the electric field normalized to the Dreicer field $E_{\rm D}$ and the effective charge~\citep{connor}. One may therefore expect that the differences between the Fokker--Planck and the Boltzmann operator are amplified in the Dreicer growth rate, which is verified in figure~\ref{fig:Dreicer}. 
Most notably, the partially screened collision operator reduces the Dreicer growth rate by several orders of magnitude compared to the completely screened case. 
In contrast, the Fokker--Planck and the Boltzmann operator exhibit a similar qualitative behaviour, with differences around tens of percent in most of the interval. 
Although significant, this growth rate difference between the two collision operators is small compared to uncertainties in both experimental parameters and the collision operator. As discussed in Sec.~\ref{sec:collop}, the latter is because the validity of the Born approximation breaks down at the low critical momenta obtained with the electric fields in figure~\ref{fig:Dreicer}. 
Consequently, the differences between the Fokker--Planck and the Boltzmann operator can not be regarded as practically relevant. 

\section{Conclusions}
\label{sec:concl}
Collisions between fast electrons and partially ionized atoms are sensitive to the effect of screening.  In this paper, we derived a collision operator accounting for the effect of partial screening. This generalization of the Fokker--Planck operator in a fully ionized plasma can be expressed as modifications to the deflection and slowing-down frequencies.  
To obtain these collision frequencies, we treated the interaction between fast electrons and partially ionized impurities quantum-mechanically in the Born approximation.
We used DFT calculations to obtain the electron density
distribution of the impurity ions, which determined the differential
cross sections for elastic scattering. This allowed us
to define an effective ion length scale, and we display these results
in table~\ref{tab:consts} for the ion species that are most common in
fusion experiments: helium, beryllium, carbon, nitrogen, neon, argon,
xenon and tungsten. The results showed that a formula for this length scale  based on the
Thomas-Fermi model usually suffices for an accurate description of
screening effects. However, the length scales derived from DFT give
higher accuracy, especially for low electron momenta. Combined with a stopping-power description of inelastic scattering, this forms the generalized collision operator for fast electrons interacting with partially ionized impurities.

Using the generalized collision operator, the runaway growth rate and
energy spectrum were calculated. 
Unlike the completely screened description, screening effects lead to a stronger-than-linear electric-field dependence causing a significantly enhanced avalanche growth rate at high electric fields. 
This behaviour contrasts previous results~\citep{Putvinski1997}, which predicted the growth rate to always be reduced compared to the completely screened limit.    
At weak electric fields, partial screening however reduces the avalanche growth rate by significantly enhancing the threshold field.
In addition, we found that the exponentially decaying avalanche-dominated energy spectrum has an average energy that depends on the electric field. This energy is significantly lower than with complete screening, which is equivalent to a fully ionized plasma having the same effective charge.

Finally, we showed that 
the validity of the Fokker--Planck equation is less clearly satisfied  
for partially screened collisions than in the pure Coulomb case, 
 due to the enhancement of large momentum transfers. Despite this, we found that 
the runaway energy and growth rate
are well captured by a treatment based on the Fokker-Planck
operator. The overall shape of the fast electron distribution is
somewhat different in the more precise Boltzmann approach, but this
has negligible effect on the integrated quantities such as the energy
spectrum and runaway current. However, quantities which are highly sensitive to the angular distribution, such as synchrotron radiation, can be moderately affected in high-electric-field cases.

\acknowledgments
The authors are grateful to S Newton, G Wilkie and I Pusztai for fruitful
discussions.  This work was supported by the Swedish Research Council
(Dnr.~2014-5510), the Knut and Alice Wallenberg Foundation and the
European Research Council (ERC-2014-CoG grant 647121).  The work has
been carried out within the framework of the EUROfusion Consortium and
has received funding from the Euratom research and training programme
2014-2018 under grant agreement No 633053. The views and opinions
expressed herein do not necessarily reflect those of the European
Commission.
\appendix
\section{Evaluating the terms in the collision operator with covariant notation}
\label{app:covariant}
To obtain an explicit form of the collision operator in spherical coordinates $\{p, \theta, \phi\}$ where $\mathbf{p} = (p,0,0)$, we transform the expressions in equation~\eqref{eq:<Deltap1>v2} into an arbitrary coordinate system $\{ {\mathbf e}^\mu\}$, where the moments are
\begin{equation}\label{eq:deltas}
\begin{aligned}
\left\langle\Delta p^\mu \right\rangle_{{\rm e}j}&= (\mathbf{e}^\mu \cdot \mathbf{e}_{L,j})\Delta p_L^\nu  \\&= \frac{p^\mu}{p}\left\langle \Delta p_L^1\right\rangle\,,  \\
\left\langle\Delta p^\mu \Delta p^\nu\right\rangle_{{\rm e}j}&=(\mathbf{e}^\mu \cdot \mathbf{e}_{L,\rho})(\mathbf{e}^\nu \cdot \mathbf{e}_{L,\sigma})\Delta u_L^\rho\Delta u_L^l \\ &= \left[\delta^{\mu\nu} - \frac{p^\mu p^\nu}{p^2}\right]\left\langle \Delta p_L^2 \Delta p_L^2 \right\rangle\,.
\end{aligned}
\end{equation}

We now wish to convert the expressions~\eqref{eq:deltas} into the coordinate basis $\{p, \theta, \phi\}$. In this system, the three-dimensional metric  is 
\begin{equation}
g_{\mu\nu} = 
\begin{pmatrix}
1 & 0 & 0 \\
0 & p^2 & 0 \\
0 & 0 & p^2 \sin^2\theta
\end{pmatrix}.
\end{equation}
Note that to convert the expressions in equation~\eqref{eq:deltas} from a normalized basis into a coordinate basis, any contravector $V^\mu$ must be multiplied by a factor of the square root of the inverse metric: $``\sqrt{g^{\mu\mu}}" = [1,1/p,1/(p \sin \theta)]^\mu$ and similarly for tensors. In covariant notation, the divergence can be written elegantly as 
\begin{equation}
\nabla_\mu V^\mu = \frac{1}{\sqrt{g}} \partial_\mu(\sqrt{g}   V^\mu),
\end{equation}
where $\sqrt{g} = \sqrt{|\text{det}(g_{\mu\nu})|} =p^2 \sin\theta$, while the second-order differential operator in the Fokker--Planck terms requires Christoffel symbols 
$\Gamma^{\rho}_{\mu\nu} = \frac{1}{2}g^{\rho \sigma}\left( \partial_\nu g_{\sigma \mu} + \partial_\mu g_{\sigma \nu} - \partial_\sigma g_{\mu\nu}\right),$
according to 
\begin{equation}
\nabla_\nu T^{\mu\nu} = \partial_\nu  T^{\mu\nu} + \Gamma^{\mu}_{\nu\rho}T^{\rho\nu} +
 \Gamma^{\nu}_{\nu\rho}T^{\mu\rho}.
\end{equation}
Thus,
\begin{align}
 C^{{\rm e}j} &=\frac{1}{\sqrt{g}} \partial_\mu
\left( \sqrt{g}  V^\mu 
\right),\label{eq:ceiIs} \\
V^\mu &= 
 -f_{\rm e} \left\langle\Delta p^\mu \right\rangle_{{\rm e}j}
 +\frac{1}{2}\left[
 \partial_\nu (f_{\rm e} \left\langle \Delta p^\mu \Delta p^\nu \right\rangle_{{\rm e}j})
 +
 \Gamma^{\mu}_{\nu\rho} (f_{\rm e} \left\langle \Delta p^\rho \Delta p^\nu \right\rangle_{{\rm e}j}) +
 \Gamma^{\nu}_{\nu\rho}
 f_{\rm e} \left\langle \Delta p^\mu \Delta p^\rho \right\rangle_{{\rm e}j}\right],
\end{align} 
and $\Gamma^{\rho}_{\mu\nu}$ has the following non-zero components:
\begin{eqnarray}
&\Gamma^{1}_{2 2} = -p , \quad
&\Gamma^{1}_{3 3} = -p \sin^2\theta, \\
&\Gamma^{2}_{2 1} = 1/p,\quad
&\Gamma^{2}_{3 3} = -\cos\theta \sin\theta, \\
&\Gamma^{3}_{3 1} = 1/p,\quad
&\Gamma^{3}_{3 2} = \cot\theta.
\end{eqnarray}
This yields
\begin{align}
V^1 &= 
 -
 \left[\left\langle\Delta p_L^1 \right\rangle_{{\rm e}j}+ 
 \frac{1}{p } \left\langle \Delta p_L^2 \Delta p_L^2\right\rangle_{{\rm e}j}
 \right] f_{\rm e} =0 \,,
\\ 
%----------------------------------
V^2 &= 
 \frac{1}{2 p^2} \left\langle \Delta p_L^2 \Delta p_L^2\right\rangle_{{\rm e}j}
\partial_\theta f_{\rm e} \,,
\\ 
%--------------------------------
V^{3} &= 
 \frac{1}{2 p^2\sin^2 \theta} 
  \left\langle \Delta p_L^2 \Delta p_L^2\right\rangle_{{\rm e}j}\partial_\phi f_{\rm e}\,.
\end{align}

\section{General properties of the screening function: high-energy behaviour}

\label{app:gHighEnergy}
Utilizing the fact that $F_j(q) \rightarrow 0$ for $q \gg 1$ and $F_j(q) \rightarrow N_{{\rm e},j}$ for $q \ll 1$, we can find a closed expression for $g_j(p)$ in the limit of large $y = 2 p/\alpha  = q/x$ which is then valid from mildly relativistic energies (if the transition from complete screening to full screening in the form factor is located around $y\sim 1 \Leftrightarrow p \sim 10^{-2}$). 
The screening function is defined as
\begin{align}
g_j(p) &= \int_{1/\Lambda}^1  \left[\left|Z_j-F_j(q)\right|^2 - Z_{0,j}^2\right] \frac{\dd x}{x} \nonumber \\
&\approx \lim_{\Lambda\rightarrow \infty}\int_{y/\Lambda}^y \left\{2Z_j\left[N_{{\rm e},j}-F_j(q)\right] + F_j^2(q)-N_{{\rm e},j}^2\right\} \frac{\dd q}{q}, 
\label{eq:g2}
\end{align}
 For simplicity, we normalize the radial coordinate to the Bohr radius $a_0$ and the density such that $N_{{\rm e},j} = 4 \pi\int r^2 \rho_{{\rm e},j}( r) \dd  r$. The form factor (for a spherically averaged charge distribution) is then determined by 
\begin{equation}
F_j(q) =  
4 \pi \int_0^\infty  \rho_{{\rm e},j}( r) \frac{ r }{ q}\sin(q  r)\,\dd  r\,,
\end{equation}
 
The first term of equation~\eqref{eq:g2} can be simplified using partial integration, and extending the remaining integral to infinity:
\begin{align}
I_{1,j} \equiv& \ 2Z_j \int_{y/\Lambda}^y  \left[N_{{\rm e},j}-F_j(q)\right] \ \frac{\dd q}{q}  
\nonumber\\
=&\ 2Z_j \left(\left[\vphantom{\frac{1}{2}} \ln q \left[N_{{\rm e},j}-F_j(q)\right]\right]_{y/\Lambda}^y - \int_0^\infty \!\ln q \, F_j'(q) \dd q \right).
\end{align}
Note that if the atom has a spherically symmetric potential, the mean dipole moment ($\propto \int d^3 r  \,\mathbf{r} n(\mathbf{r})$) vanishes~\citep{landauQM}, in which case the first derivative of the form factor vanishes identically for small arguments.
Utilizing this fact for $F(y/\Lambda \ll 1) = N_{{\rm e},j}$ and $F_j(y \gg 1) = 0$, we obtain
\begin{align}
 I_{1,j} =&\
2Z_j N_{{\rm e},j} \ln y 
+
 8 Z_j \pi \int_0^\infty \rho_{{\rm e},j}(r) r^2 \dd r \!\underbrace{ \int_0^\infty \frac{\ln q}{q} 
\left(
\cos (qr)
-\frac{\sin(qr) }{r q}
\right)\dd q}_{=\gamma_E-1+\ln r}
\nonumber\\
=&\ 2Z_j N_{{\rm e},j} \left(\ln y - 1 + \gamma_E + \hat I_{1,j} \right),
\end{align}
where we used $4 \pi \int r^2 \rho_{{\rm e},j}(r) \dd r=N_{{\rm e},j}$ and
\begin{equation}
\hat I_{1,j} \equiv \frac{4 \pi}{N_{{\rm e},j}} \int_0^\infty \!\rho_{{\rm e},j}(r) r^2 \ln r \dd r \label{eq:Ihat1}.
\end{equation}

Similarly, for the second term,
\begin{align}
I_{2,j} \equiv &\ \int_{1/\Lambda}^1\! \left\{ F_j^2(q)-N_{{\rm e},j}^2\right\} \frac{\dd x}{x}
\nonumber\\=&\  
\left[\vphantom{\frac{1}{2}} \ln q \left[F_j(q)^2 - N_{{\rm e},j}^2\right]\right]_{y/\Lambda}^y \!\!- 2\int_0^\infty \ln q F_j(q) F_j'(q) \dd q 
\nonumber \\ =&\
%%%%%%%%%%%%%%
- N_{{\rm e},j}^2 \ln y 
- (4 \pi)^2 \!\int_0^\infty \rho_{{\rm e},j}(r)r^2 \dd r \int_0^\infty \rho_{{\rm e},j}(r_2) r_2^2 \dd r_2  
\int_0^\infty 2\frac{\ln q}{q} \frac{\sin (q r_2)}{q r_2} \left(\cos(q r) - \frac{\sin(q r)}{q r} \right)
\nonumber \\ =&\
- N_{{\rm e},j}^2 \ln y - (4 \pi)^2 \!\int_0^\infty \rho_{{\rm e},j}(r)r^2 \dd r \int_0^\infty \rho_{{\rm e},j}(r_2) r_2^2 \dd r_2 \nonumber\\&
\quad \times \left[\gamma_E-\frac{3}{2}
+
\frac{(r+r_2)^2}{4 r r_2} \ln(r+r_2) - \frac{(r-r_2)^2}{4 r r_2}\ln|r-r_2|
\right. \nonumber\\& \quad\left.
+\frac{(r^2-r_2^2)}{4 r r_2}\ln\!\left(\frac{r+r_2}{|r-r_2|}\right)\left[\ln\left(r^2-r_2^2\right)+2(\gamma_E-1)\right]
 \right].
 \end{align}
In the integrand, the first term is straightforward to integrate with $4 \pi \int r^2 \rho_{{\rm e},j}(r) \dd r=N_{{\rm e},j}$, while the last term must vanish upon integration since it is antisymmetric in $r-r_2$, leaving  
 \begin{align}
 %%%%%%%%%%%%%%
I_{2,j} =&\
 -N_{{\rm e},j}^2\left(\ln y -\frac{3}{2}+\gamma_E+ \hat I_{2,j} \right)  ,
\end{align}
where 
\begin{align}
\hat I_{2,j} \equiv&\  \frac{(4 \pi)^2}{4 N_{{\rm e},j}^2}\! \!\int_0^\infty\int_0^\infty \rho_{{\rm e},j}(r)r   \rho_{{\rm e},j}(r_2) r_2  % 
\left[
(r+r_2)^2 \ln(r+r_2) - (r-r_2)^2\ln|r-r_2|%\vphantom{\frac{2}{2}}
\right]\dd r_2 \dd r
\nonumber\\=&\
 \frac{(4 \pi)^2}{16 N_{{\rm e},j}^2}\! \!\int_0^\infty \!\! \dd s \int_0^s \!\!\dd t \, 
(s^2-t^2) \rho_{{\rm e},j}\Big(\frac{s+t}{2}\Big)
 \rho_{{\rm e},j}\Big(\frac{s-t}{2}\Big)  
\left[
s^2 \ln s - t^2\ln t
\right].
\label{eq:Ihat2}
\end{align}

Adding the terms of equation~\eqref{eq:g2} together yields (using $2 ZN_{\rm e}-N_{\rm e}^2 = Z^2-Z_0^2$)
\begin{align}
g_j(p) =&\ I_{1,j} + I_{2,j}  
\nonumber\\ =&\ (Z_j^2-Z_{0,j}^2)[ \ln \left(2 p/\alpha \right) -1+\gamma_E ] 
+ 2 Z_j N_{{\rm e},j} \hat I_{1,j}
 + 
N_{{\rm e},j}^2\Big(\frac{1}{2}-\hat I_{2,j} \Big) .
\label{eq:gGeneralBehavior}
\end{align}
Hence, the screening function $g_j(p)$ grows logarithmically with momentum at high electron energies. This allows us to determine $a_j$ so that the deflection frequency exactly matches the high-energy asymptote of the DFT results. 
 Matching equation~\eqref{eq:gGeneralBehavior} with the high-energy asymptote of $g_j(p)$ from equation~\eqref{eq:g}, 
\begin{equation}
g_j(p) \sim (Z_j^2-Z_{0,j}^2)\ln (p \bar a_j) -\frac{2}{3}N_{{\rm e},j}^2, \qquad p \bar{a}_j \gg 1,
\label{eq:gAsymptote}
\end{equation}
we obtain
\begin{align}
\label{eq:abar} \bar a_j = \frac{2}{\alpha}\exp \left[\gamma_E-1 + \frac{2 Z_j\hat I_{1,j}+ N_{{\rm e},j}\big(\nofrac{7}{6} - \hat I_{2,j} \big)}{Z_j+Z_{0,j}}\right].
\end{align}
The values of $\bar a_j $ are given for many of the fusion-relevant ion species in table~\ref{tab:consts}, of which the constants for argon and neon are illustrated in figure~\ref{fig:a} as a function of $Z_0$ in solid line. 

\begin{table}
\centering
\begin{tabular}{llcllcllcllc}
Ion&$\bar a_j$&\phantom{a}&Ion&$\bar a_j$&\phantom{a}&Ion&$\bar a_j$&\phantom{a}&Ion&$\bar a_j$ \\ \cline{1-2}\cline{4-5}\cline{7-8}\cline{10-11} \tallrow He$^{0}$&173&&N$^{0}$&135&&Ar$^{0}$&96&&Xe$^{1+}$&65&\\ \tallrow
He$^{1+}$&123&&N$^{1+}$&115&&Ar$^{1+}$&90&&Xe$^{2+}$&63&\\ \tallrow
Be$^{0}$&159&&N$^{2+}$&97&&Ar$^{2+}$&84&&Xe$^{3+}$&61&\\ \tallrow
Be$^{1+}$&114&&N$^{3+}$&79&&Ar$^{3+}$&78&&W$^{0}$&59&\\ \tallrow
Be$^{2+}$&67&&N$^{4+}$&59&&Ar$^{4+}$&72&&W$^{30+}$&33&\\ \tallrow
Be$^{3+}$&59&&N$^{5+}$&35&&Ar$^{5+}$&65&&W$^{40+}$&25&\\ \tallrow
C$^{0}$&144&&N$^{6+}$&33&&Ar$^{6+}$&59&&W$^{50+}$&18&\\ \tallrow
C$^{1+}$&118&&Ne$^{0}$&111&&Ar$^{7+}$&53&&W$^{60+}$&13&\\ \tallrow
C$^{2+}$&95&&Ne$^{1+}$&100&&Ar$^{8+}$&47&&& &\\ \tallrow
C$^{3+}$&70&&Ne$^{2+}$&90&&Ar$^{9+}$&44&&& &\\ \tallrow
C$^{4+}$&42&&Ne$^{3+}$&80&&Ar$^{10+}$&41&&& &\\ \tallrow
C$^{5+}$&39&&Ne$^{4+}$&71&&Ar$^{11+}$&38&&& &\\ \tallrow
& &&Ne$^{5+}$&62&&Ar$^{12+}$&35&&& &\\ \tallrow
& &&Ne$^{6+}$&52&&Ar$^{13+}$&32&&& &\\ \tallrow
& &&Ne$^{7+}$&40&&Ar$^{14+}$&27&&& &\\ \tallrow
& &&Ne$^{8+}$&24&&Ar$^{15+}$&21&&& &\\ \tallrow
& &&Ne$^{9+}$&23&&Ar$^{16+}$&13&&& &\\ \tallrow
& &&& &&Ar$^{17+}$&13&&& &\\ \tallrow
\end{tabular}
\caption{\label{tab:consts}Values of the normalized effective length scale $\bar a_j = 2 a_j/\alpha$ for different ion species. These values were obtained with equation~\eqref{eq:abar} using electronic charge densities from DFT calculations.}
\end{table}

\section{Partially screened avalanche-dominated runaway energy spectrum}
\label{app:spectrum}
We here generalize the derivation of the high electric field, avalanche-dominated distribution by~\citet{fulop2006} to account for partially ionized impurities. In \citet{fulop2006}, the kinetic equation is specialized to the case where $E \gg E_c$, which gives a narrow pitch-angle distribution where the majority of the runaway electrons populate the region $1-\xi \ll 1$, which is used as an expansion parameter.  Note however, that assuming fast pitch-angle dynamics \citep{lehtinen, aleynikovPRL} is invalid when $E \gg \Eceffeq$, where $\Eceffeq$ is the effective critical field~\citep{Ecrit}.

Neglecting how the avalanche source term affects the shape of the distribution, we solve the coupled equations given by the avalanche growth rate~\eqref{eq:Gamma}
and the kinetic equation. In the kinetic equation, we  utilize $E \gg \Eceffeq$ to replace the friction terms by \Eceff\ in order to match the near-critical behaviour~\citep{Ecrit}:
\begin{align}
\tau_c \frac{\partial \bar{f}}{\partial t} &= \frac{\partial}{\partial p} 
\Bigg[\Bigg(- \frac{\xi E}{E_{\rm c}} + \underbrace{p\nu_{\rm s}+ \Fbr+ \frac{p\gamma}{\taur} (1-\xi^2)}_{
\sim  \Eceffeq/E_{\rm c}}\Bigg)\bar{f} \Bigg] \nonumber \\ 
&\quad+\frac{\partial}{\partial \xi}\bigg[(1-\xi^2)\left(-\frac{1}{p}\frac{E}{E_{\rm c}}\bar{f}+\frac{1}{2}\nu_{\rm D}\frac{\partial \bar{f}}{\partial \xi}\right) -\underbrace{\frac{\xi(1-\xi^2)}{\taur \gamma}}_{\rm neglect}\bar{f}\bigg] \, \label{eq:FP2} 
\end{align}   
Here, $\bar f = p^2 f$, $\Fbr$ describes bremsstrahlung losses and $\taur$ is a measure of the synchrotron losses. Assuming that the distribution is narrow, $p_\perp \ll p_\parallel \simeq p$, so that $1-\xi \ll 1$, we integrate equation~\eqref{eq:FP2} over $\xi$. Together with equation~\eqref{eq:Gamma}, we obtain
  \begin{align}
\tau_c \Gamma (E)
F
 + \frac{E-\Eceffeq}{E_c}\frac{\p F}{\p p} =0,
 \end{align}
which has the solution
  \begin{equation}
  F(p,t) = n_{\rm RE}(t) \frac{1}{p_0}\mathrm{e}^{-p/p_0},
  \end{equation}
where 
    $$p_0 =   \frac{ E-\Eceffeq}{E_c \tau_c \Gamma(E)}  = \frac{e}{m_{\rm e} c}\frac{  E-\Eceffeq}{ \Gamma(E) }. $$
Since $\Gamma \propto E-\Eceffeq$ for $E/\Eceffeq-1 \ll 1$, the term $\Eceffeq$ ensures that $  p_0 < \infty$ in the limit $E \rightarrow \Eceffeq$.
The average runaway momentum $p_0$ can alternatively be interpreted as an average energy since $p_0\gg1$ typically. Although $p_0$ only depends on the effective charge in the fully ionized case, the average momentum acquires a significant $E$-dependence in the presence of partially screened ions, as shown in figure~\ref{fig:ava}.

\bibliographystyle{jpp}

\bibliography{references} % references.bib
\end{document}